\documentclass[aps, prd, preprint, a4paper, superscriptaddress, nofootinbib]{revtex4-1}
\usepackage[dvipdfmx]{graphicx}
\usepackage{amsmath,amssymb}
\usepackage{comment}
\usepackage{here}
\usepackage{docmute}
\usepackage{simplewick}
\usepackage{hyperref}
\usepackage{layout}
\usepackage[dvipdfmx]{color}
\usepackage{physics}

\usepackage[normalem]{ulem} 
\usepackage{cancel}
\newcommand{\A}{\mathcal{A}}
  \newcommand{\B}{\mathcal{B}}
  
  \newcommand{\D}{\mathcal{D}}

  \renewcommand{\H}{\mathcal{H}}
  
  \newcommand{\J}{\mathcal{J}}
  \newcommand{\K}{\mathcal{K}}

  \renewcommand{\O}{\mathcal{O}}

  \newcommand{\R}{\mathcal{R}}

  \newcommand{\V}{\mathcal{V}}

 \def \be {\begin{equation}}
\def \ee {\end{equation}}
\def \beqa {\begin{eqnarray}}
\def \eeqa {\end{eqnarray}}
\def \nn {\notag}
\begin{document}

\title{ Lattice study on a tetra-quark state $T_{bb}$ in the HAL QCD method}

\author{Takafumi Aoki}
\email{takafumi.aoki@yukawa.kyoto-u.ac.jp}
\affiliation{Yukawa Institute for Theoretical Physics\\Kyoto University, Kyoto 606-8502, Japan}
\author{Sinya Aoki}
\email{saoki@yukawa.kyoto-u.ac.jp }
\affiliation{Center for Gravitational Physics and Quantum Information, Yukawa Institute for Theoretical Physics, Kyoto University, Kyoto 606-8502, Japan}
\affiliation{RIKEN Nishina Center (RNC), Saitama 351-0198, Japan}
\author{Takashi Inoue}
\email{inoue.takashi@nihon-u.ac.jp }
\affiliation{Nihon University, College of Bioresource Sciences, Kanagawa 252-0880, Japan}

\preprint{YITP-22-117} 
\begin{abstract}
We study a doubly-bottomed tetra-quark state $(bb\bar{u}\bar{d})$ with quantum number $I(J^P)=0(1^+)$, 
denoted by $T_{bb}$, in lattice QCD  with the Non-Relativistic QCD (NRQCD) quark action for $b$ quarks.
Employing $(2+1)$-flavor gauge configurations at $a \approx 0.09$ {fm} on $32^3\times 64$ lattices,
we have extracted the coupled channel potential between $\bar{B}\bar{B}^*$ and $\bar{B}^* \bar{B}^*$ in the  HAL QCD method,
which predicts an existence of a bound  $T_{bb}$ below the $\bar{B}\bar{B}^*$ threshold.
By extrapolating results at $m_\pi\approx 410,\, 570,\, 700$ {MeV} to the physical pion mass $m_\pi\approx140$ {MeV},
we obtain a biding energy with its statistical error as $E_{\rm binding}^{\rm (single)} = 155(17)$ MeV and $E_{\rm binding}^{\rm (coupled)} = 83(10)$ MeV, where ``coupled" means that effects due to virtual $\bar{B}^* \bar{B}^*$ states are included through the coupled channel potential, while
only a potential for a single $\bar{B}\bar{B}^*$ channel is used in the analysis for ``single".
A comparison shows  that the effect from virtual $\bar{B}^* \bar{B}^*$ states is quite sizable to the binding energy of $T_{bb}$. 
We estimate systematic errors to be $\pm 20$ MeV at most, which are mainly caused by the NRQCD approximation for $b$ quarks. 
\end{abstract}
\maketitle

\section{Introduction} \label{sect:intro}

One of typical characteristic features of QCD is the color confinement that only color-singlet states can appear in Nature.
While almost all observed color-singlets states are either mesons ($q\bar{q}$) or baryons ($qqq$), 
other color-singlet states such as tetra-quark state ($qq\bar{q}\bar{q}$), penta-quark states ($qqqq\bar{q}$) and 
glueball states are theoretically allowed to exist.
These states are rarely observed and called exotic hadrons, whose existences have not been firmly established yet.
Recently experimental observations have been reported for several heavy exotic hadrons, which include
tetra-quark states $X(2900)$\cite{LHCb:2020pxc} and $T_{cc}$\cite{LHCb:2021vvq} containing  one or two charm quarks,  
a penta-quark state $P_c$\cite{LHCb:2015yax} containing a charm and anti-charm pair, or 
tetra-quark states $Z_b$\cite{Belle:2011aa} containing a bottom and anti-bottom pair.
Their properties such as particle contents and internal structures, however, are needed to be understood, in particular, theoretically
in terms of QCD.

In this paper, as the first  step to understand such heavy exotic hadrons, 
we investigate a tetra-quark hadron $(bb\bar{u}\bar{d})$ in $I(J^P)=0(1^+)$ channel, called $T_{bb}$,
from the first principle using lattice QCD.
While $T_{bb}$ has not been experimentally observed yet, 
theoretical predictions by the di-quark model\cite{Cheng:2020wxa} and by color magnetic interactions under the static limit\cite{Karliner:2017qjm,Eichten:2017ffp}
suggest existences of  heavy tetra-quark bound states $QQ\bar{q}\bar{q}$.
Indeed, as mentioned before,  $T_{cc}$, a charm counterpart  to $T_{bb}$, seems to exist.

There exist several lattice QCD studies for  $T_{bb}$\cite{Bicudo:2015kna,Bicudo:2016ooe,Francis:2016hui,Junnarkar:2018twb,Leskovec:2019ioa,Mohanta:2020eed,Hudspith:2023loy},
all of which conclude that $T_{bb}$  appears as a bound state below the $\bar{B} \bar{B}^*$ threshold,
where the threshold energy is given by $E^{\rm threshold}_{\bar{B} \bar{B}^*} \simeq 10.604 $ GeV.
A predicted binding energy, however, depends on both a treatment of heavy $b$ quarks and a number of channels included in lattice calculations,
as shown in Tab.~\ref{tab:prev_binding_energy}.
Within a static quark approximation for $b$ quarks, the binding energy is about 30 MeV larger in a single $\B$ channel analysis
than in a coupled  $\B$ and $\B^*$ channel analysis, where the $\bar{B} \bar{B}^*$ channel is denoted by $\B$
while the $\bar{B}^* \bar{B}^*$ channel by $\B^*$, whose threshold energy $E^{\rm threshold}_{\bar{B}^* \bar{B}^*} \simeq
10.649$ GeV is about 45 MeV above the $\bar{B} \bar{B}^*$ threshold.
Within a single channel analysis,  the binding energy increases by about 40 MeV,
 if  $b$ quarks are treated by the lattice NRQCD, which allows $b$ quarks to move not only in time but also in space.
Then one may ask what happens for a combination of NRQCD $b$ quarks and the coupled channel analysis.

Unlike the static quark approximation where the coupled channel analysis can be applied to the Shr\"odinger equation 
with static quark potentials, 
the coupled channel analysis can not be directly applied to the finite volume method with moving $b$ quarks in the NRQCD,
where only energies of states can be obtained.
Above $\bar{B}^* \bar{B}^*$ threshold, coupled channel treatments become possible
by the finite volume method\cite{Luscher:1990ux},  
which however requires some assumptions on the coupled channel $S$-matrix.
  
 In this paper, to calculate the binding energy of $T_{bb}$ by the coupled channel analysis with the NRQCD action for $b$ quarks,
 we alternatively employ the couple channel extension of  the HAL QCD method\cite{Aoki:2013cra},
 by which we can extract the coupled channel potential directly.
 We review the HAL QCD method in Sec.~\ref{sec:HALQCD} and summarize our lattice QCD setup,
 including the NRQCD action for  $b$ quarks in Sec.~\ref{sec:LQCD_setup}. 
 In Sec.~\ref{sec:numerical_result}, we present results of potentials and the scattering analysis.
 Finally we give a summary of  this study in Sec.~\ref{sec:conclusion}.

\begin{table}[h]
  \centering
  \begin{tabular}{c|c|c}
     $b$ quark\textbackslash channels & $\B(\equiv \bar{B}\bar{B}^*)$ & $\B$ and $\B^*(\equiv \bar{B}^*\bar{B}^*)$ \\
    \hline
      Static approx& $90(^{+43}_{-36})$ {MeV}\cite{Bicudo:2015kna} & $59(^{+30}_{-38})$ {MeV}\cite{Bicudo:2016ooe} \\
      NRQCD approx& $180(10)(3)$ {MeV}\cite{Francis:2016hui} &  \\
                                 & $165(33)$ {MeV}\cite{Junnarkar:2018twb} & \\ 
                                & $128(24)(10)$ {MeV}\cite{Leskovec:2019ioa} & \\ 
                                 & $186(22)$ {MeV}\cite{Mohanta:2020eed} & \\ 
                                 & $112(13)$ {MeV}\cite{Hudspith:2023loy} & \\
    \hline
  \end{tabular}
  \caption{Binding energies extrapolated to the physical pion mass in previous lattice studies. 
  \label{tab:prev_binding_energy}}
\end{table}

\section{HAL QCD method}
\label{sec:HALQCD}

\subsection{Definition of the Potential}
A basic quantity for a definition of potentials in the HAL QCD method is  the Euclidean time Nambu-Bethe-Salpeter(NBS) wave function,
defined by\cite{Ishii:2006ec,Aoki:2009ji,Aoki:2012tk,Aoki:2013cra} 
\beqa
  \psi_W^{H_1+H_2}({\bf r},t) \equiv \psi^{H_1+H_2}_{W}({\bf r})\, e^{-Wt}
  &\equiv& \frac{1}{\sqrt{Z_{H_1}}}\frac{1}{\sqrt{Z_{H_2}}}\sum_{\bf x}\bra{\Omega} H_1({\bf x}+{\bf r},t)H_2({\bf x},t) \ket{(H_1+H_2);W},\nn \\
\eeqa
where $H_i({\bf x},t)$ is the hadron operator at $({\bf x},t)$,  
$\ket{\Omega}$ is the QCD vacuum state, 
$\ket*{(H_1+H_2);W}$ stands for an eigenstate of the QCD Hamiltonian having quantum numbers of the two-hadrons $H_1+H_2$ with a center-of-mass energy $W$, and 
$Z_{H_i}=\vert \bra{\Omega} H_i(0) \ket{H_i}\vert^2$ with $\ket{H_i}$ being a single hadron state.
We focus our attention on an energy region below inelastic threshold, where only elastic scattering occurs.
In this energy region,
the asymptotic behavior of the $\ell$-th partial-wave of the NBS wave function reads
\beqa
  \label{eq:NBS_behaviar}
  \psi^{H_1+H_2}_{W,\ell}({\bf r}) &\xrightarrow{r\to\infty} & 
   \left[j_\ell(p_W^{}r)-\pi t_\ell(W)h_\ell^+(p_W^{}r) \right]P_\ell(\hat{\vb{r}}\cdot\hat{{\bf p}}_W^{}) , 
\eeqa
where a magnitude of a relative momentum $p_W^{}$ is determined from a relation $W=E_{W1}+E_{W2}=\sqrt{p^2_W+m_{H_1}^2}+\sqrt{p_W^2+m_{H_2}^2}$,  $P_\ell(z)$ is the Legendre polynomial, $j_\ell(z)\slash n_\ell(z)$ is the spherical Bessel$\slash$Neumann function, and $h_\ell^\pm(z)=n_\ell(z)\pm i j_\ell(z)$ are spherical Hankel functions.
The scattering $T$-matrix $t_\ell(W)$ in the above is related to the unitary $S$-matrix as $s_\ell(W)=1-2\pi i t_\ell(W)$, 
and to the scattering amplitude as $f_\ell(W)=-\frac{\pi}{p_W^{}}t_\ell(W)$. 

A hadronic 4-point correlation function in lattice QCD  can be expressed in terms of NBS wave functions as
\beqa
  F^{H_1+H_2}_\J({\bf r},t)
 & \equiv& \sum_{{\bf x}}\bra{\Omega}{H_1(\vb{x+r},t)H_2(\vb{x},t)\J^\dag_{H_1+H_2}(t=0)}\ket{\Omega}\nn \\
  &=& \sum_{{\bf x}}\sum_{n}\bra{\Omega}{H_1(\vb{x+r},t)H_2(\vb{x},t)}\ket{(H_1+H_2);W_n}\bra{(H_1+H_2);W_n}{\J^\dag_{H_1+H_2}(0)}\ket{\Omega}
  \nn \\ &+&
  (\mathrm{inela})\nn \\
& {\simeq}&  \sum_{n}\A_{\J,n}\psi_{W_n}^{H_1+H_2}(\vb{r},t \ge t^\mathrm{(inela)})
\quad \xrightarrow{t\to\infty} \quad  \A_{\J,0}\psi_{W_0}^{H_1+H_2}(\vb{r})\,e^{-W_0 t},
 \eeqa
 where $\J^\dag_{H_1+H_2}(0)$ is a source operator which creates two-hadron states at $t=0$ with a target quantum number $I(J^P)$ of $H_1+H_2$,  (inela) represents inelastic contributions, which become negligible at $t\ge t^\mathrm{(inela)}$, 
 $W_0$ is the lowest eigen-energy of two hadrons, and
 \beqa
 \A _{\J,n}\equiv\sqrt{Z_{H_1}}\sqrt{Z_{H_2}}\bra{(H_1+H_2);W_n}{\J^\dag_{H_1+H_2}(0)}\ket{\Omega}.
\eeqa

In the HAL QCD method, 
a non-local but energy-independent potential  $U(\vb{r,r'})$ is formally defined from the NBS wave function 
so as to satisfy the Schr\"odinger equation below inelastic threshold as
\beqa
  \left(\frac{\nabla^2}{2\mu}+\frac{p_W^2}{2\mu}\right)\psi_W({\bf r}) =  \int\dd[3]{{\bf r}'}U({\bf r},{\bf r'})\psi_W({\bf r'}),
\eeqa
 where $\mu$ is the reduced mass of two hadrons.
Since QCD interactions are short-ranged, $U(\vb{r,r'})$ vanishes sufficiently fast as $\vert \vb{r}\vert$ increases.
The potential  $U(\vb{r,r'})$ may depend on how  sink hadron operators $H_1$ and $H_2$ are constructed from quarks.  
Even though a choice of hadron operators is fixed, however, the above equation can not determine $U(\vb{r,r'})$ uniquely due to a restriction of the energy  below the inelastic threshold\cite{Aoki:2011gt,Aoki:2012bb}.
Thus the above definition of the potential is rather formal.
For concreteness,  
we define $U (\vb{r,r'})$ in the derivative expansion, which is symbolically written as 
\beqa
  \label{eq:diff_expansion}
  U({\bf r},{\bf r'})&=& V({\bf r},\nabla)\delta({\bf r}-{\bf r'})= \sum_{k=0}^\infty V^{(k)}({\bf r})\nabla^{k}\delta({\bf r}-{\bf r'}),
\eeqa
and determine coefficient functions $V^{(k)}({\bf r})$ order by order.
For example,  the leading order term can be approximately obtained as
\beqa
  V^{(0)}({\bf r};W)= \frac{1}{\psi_W({\bf r})}\qty(\frac{\nabla^2}{2\mu}+\frac{p_W^2}{2\mu})\psi_W({\bf r}),
\eeqa
where  $V^{(0)}({\bf r};W)$, obtained from the NBS wave function $\psi_W({\bf r})$,  is the leading order approximation of $V^{(0)}({\bf r})$.
Given the relationship between the hadron 4-point correlation function and the NBS wave function, the LO potential from the ground state  is extracted  as
\beqa
  V^{(0)}({\bf r}; W_0)\simeq \frac{1}{F_\J^{H_1+H_2}({\bf r},t)}\qty(\frac{\nabla^2}{2\mu}+\frac{p_{W_0}^2}{2\mu})F_\J^{H_1+H_2}({\bf r},t),
  \label{eq:original}
\eeqa
where $t$ should be taken as large as possible to make the lowest energy state dominate in the 4-point correlation function.

\subsection{Time-Dependent Method}
\label{sec:t-dep}
In order to achieve the ground state saturation in eq.~\eqref{eq:original}, $t$ should satisfy $ t \gg 1/(W_1-W_0) \propto L^2$  for two hadron systems, where $L$ is a size of the spatial extension.
Since the 4-point function $F_\J^{H_1+H_2}({\bf r},t)$ becomes very noisy at such large $t$, in particular for two baryon systems,  
it is impractical to employ eq.~\eqref{eq:original} for reliable  extractions of potentials.
An improved method of extracting the potential that does not require the ground state saturation has been proposed in Ref.~\cite{Ishii:2012ssm}, and is employed in this study.

 In the improved method, the potential can be extracted directly from a normalized 4-point function,  
called a $R$-correlator, which is a sum of NBS wave functions as
\beqa
  R^{H_1+H_2}_\J({\bf r},t)
  &\equiv& \frac{F^{H_1+H_2}_\J({\bf r},t)}{ e^{-m_{H_1}t} e^{-m_{H_2}t}}
  \simeq \sum_n \A_{\J,n}\psi_{W_n}^{H_1+H_2}({\bf r})\,e^{-\Delta W_n t},
\eeqa
where we take moderately large $t>t_\mathrm{threshold}^{\mathrm{(inela)}}$ in the right-hand side,
in order suppress inelastic contributions, and
$\Delta W_n\equiv W_n-m_{H_1}-m_{H_2}$ satisfies
\beqa
  \frac{p^2_n}{2\mu}&=& \Delta W_n+\frac{1+3\delta^2}{8\mu}(\Delta W_n)^2+\O((\Delta W_n)^3),\quad\quad \delta\equiv\frac{|m_{H_1}-m_{H_2}|}{m_{H_1}+m_{H_2}}\ .
\eeqa
Using this relation and taking $t>t_\mathrm{threshold}^{\mathrm{(inela)}}$, we obtain
\beqa
&&\int\dd[3]{{\bf r'}} U({\bf r},{\bf r'})  R^{H_1+H_2}_\J({\bf r'},t)\simeq 
  \sum_n\qty(\frac{\nabla^2}{2\mu}+\frac{p_n^2}{2\mu})A_n^\J\psi_{W_n}^{H_1+H_2}({\bf r})\, e^{-\Delta W_n t}\nn \\
  &\simeq& 
   \sum_n\qty(\frac{\nabla^2}{2\mu}+\Delta W_n+\frac{1+3\delta^2}{8\mu}(\Delta W_n)^2)A_n^\J\psi_{W_n}^{H_1+H_2}({\bf r})\,e^{-\Delta W_n t} \nn \\
   &=&  \qty(\frac{\nabla^2}{2\mu}-\pdv{t}+\frac{1+3\delta^2}{8\mu}\pdv[2]{t})R^{H_1+H_2}_\J({\bf r},t) ,
     \label{eq:def_potential_tdep}
\eeqa
which looks like a time-dependent Schr\"odinger equation for a non-local potential with relativistic corrections.
It is important to note that potentials can be extracted from a sum of NBS wave functions without knowing individual energy $\Delta W_n$ and
coefficient $A_n^\J$ by this method.
At the leading order in the derivative expansion,  eq.~\eqref{eq:def_potential_tdep} gives
\beqa
 V^{(0)}(\vb{r})&=& \frac{1}{R^{H_1+H_2}_\J(\vb{r},t)}\qty(\frac{\nabla^2}{2\mu}-\pdv{t}+\frac{1+3\delta^2}{8\mu}\pdv[2]{t})R^{H_1+H_2}_\J(\vb{r},t),
 \label{eq:LO_potential_tdep}
\eeqa
where a $t$-dependence in the right-hand side is canceled between numerator and denominator 
if inelastic contributions become negligible at $t>t_\mathrm{threshold}^{\mathrm{(inela)}}$. In practice, we use  the $t$-independence of $V^{(0)}({\bf r})$ as an indicator for $t>t_\mathrm{threshold}^{\mathrm{(inela)}}$ to satisfy.

\subsection{Coupled-Channel HAL QCD Method}
\label{sec:CC}
Since thresholds of $\B$($\bar{B} +\bar{B}^*$) and $\B^*$($\bar{B}^* +\bar{B}^*$) are so close,
we can not ignore an influence of the $\B^*$ channel to a potential in the $\B$ channel.
We thus decided to employ the coupled channel extension of the HAL QCD method in our study.

To explain this extension, we consider an energy region where an inelastic scattering $A+B\to C+D$ in addition to an elastic-scattering $A+B\to A+B$
occurs with  $m_{A}+m_{B}<m_{C}+m_{D}$.
The NBS wave function of the scattering channel $\alpha=0,1$ is denoted by
\beqa
 \psi^{\alpha}_{W;\beta}(\vb{r},t) &\equiv \psi^{\alpha}_{W;\beta}(\vb{r})&\,e^{-Wt}
 \equiv \frac{1}{\sqrt{Z^\alpha_1}\sqrt{Z^\alpha_2}}\sum_{\vb{x}}\matrixel*{\Omega}{H^\alpha_1(\vb{x+r},t)H^\alpha_2(\vb{x},t)}{W;\beta},
\eeqa
where $(H^0_1,H^0_2)=(A,B)$ or $(H^1_1,H^1_2)=(C,D)$, and $W$ is the center of mass energy. 
At a given energy $W$, there exists 2 independent states  with the same quantum number to $A+B$,
labeled by $\beta$, which are expanded in terms of asymptotic states as 
$\ket{W;\beta} =c^{0\beta}\ket{A+B;W} + c^{1\beta}\ket{C+D;W} +\cdots$.
Thus, as in the case of the elastic scattering, an asymptotic behavior of an $\ell$-th partial-wave of the NBS wave function reads\cite{Aoki:2011gt}
\beqa
 \psi^\alpha_{W;\beta,\ell}(\vb{r})
 & \xrightarrow{r\to\infty}&
  \sum_\gamma \qty[\delta^{\alpha\gamma} j_\ell(p_W^\alpha r)+ 
  p_W^\alpha h_\ell^+(p_W^\alpha r)  f_\ell^{\alpha\gamma}(W) ]c^{\gamma\beta} P_\ell(\hat{\vb{r}}\cdot\hat{\vb{p}}_W^{\alpha}),
  \label{eq:asym_CC}
\eeqa
where the scattering amplitude from a channel $\gamma$ to a channel $\alpha$ is defined from the $T$-matrix $t^{\alpha\gamma}_\ell$ as
\beqa
  f_\ell^{\alpha\gamma}(W)\equiv -\pi \sqrt{\frac{E^\alpha_{W1}E^\alpha_{W2}}{E^\gamma_{W1}E^\gamma_{W2}}}\sqrt{\frac{1}{p_W^\alpha p_W^\gamma}}t_\ell^{\alpha\gamma}(W)\ .
\eeqa
Since eq.~\eqref{eq:asym_CC} is identical to  an asymptotic solution to a coupled channel Schr\"odinger equation with the total energy $W$\cite{taylorQuantumTheoryNonrelativistic1972}, 
we define the coupled channel potential as
\beqa
  \qty(\frac{\nabla^2}{2\mu^\alpha}+{(p_W^\alpha)^2 \over 2\mu^\alpha} )\psi^{\alpha}_{W;\beta}(\vb{r})\equiv \sum_{\gamma}\int\dd[3]{\vb{r}'}U^{\alpha\gamma}(\vb{r,r'})\psi^{\gamma}_{W;\beta}(\vb{r'}),
 \label{eq:def_coupled_potential}
 \eeqa
 where $p_W^\alpha$ and $\mu^\alpha$ are a magnitude of the relative momentum  and a reduced mass in the channel $\alpha$, respectively.
 
 As in eq.~\eqref{eq:diff_expansion} for the single channel case, the non-local potential $U^{\alpha\beta}(\vb{r,r'})$
is defined in term of the derivative expansion, whose leading order term is given by
\beqa
 U^{\alpha\gamma}(\vb{r,r'})= V^{\alpha\gamma}(\vb{r})\delta(\vb{r-r'})+\O(\nabla).
 \eeqa
The LO potential can be approximately extracted  from two NBS wave functions 
by a matrix inversion as
\beqa
\left(
\begin{array}{cc}
    V^{00}(\vb{r}) & V^{01}(\vb{r}) \\
    V^{10}(\vb{r}) & V^{11}(\vb{r})
\end{array}
\right)
  &=&
 \left(
\begin{array}{cc} 
    K^{0}_{W_0;\beta_0}(\vb{r}) & K^{0}_{W_1;\beta_1}(\vb{r})\\
    K^{1}_{W_0;\beta_0}(\vb{r}) & K^{1}_{W_1;\beta_1}(\vb{r})
\end{array}
\right)
 \left(
\begin{array}{cc}     
    \psi_{W_0;\beta_0}^{0}(\vb{r}) & \psi_{W_1;\beta_1}^{0}(\vb{r})\\
    \psi_{W_0;\beta_0}^{1}(\vb{r}) & \psi_{W_1;\beta_1}^{1}(\vb{r})
\end{array}
\right)^{-1} ,
\label{eq:calc_coupled_LOpotential}
\eeqa
where  $K^\alpha_{W;\beta}(\vb{r})$ is given  by the left-hand side of eq.(\ref{eq:def_coupled_potential}). 
For the matrix inversion to obtain potentials, we must take two linearly independent NBS wave functions, by choosing $W$ and $\beta$ appropriately.
Note that it is not guaranteed that the coupled channel potential is Hermitian due to the approximation of the derivative expansion.

As in the case of the single channel, the coupled channel 4-point function is expressed in terms of NBS wave functions as
\beqa
  F^{\alpha}_\xi(\vb{r},t)
  &=& \sum_{\vb{x}}\matrixel*{\Omega}{H^\alpha_1(\vb{x+r},t)H^\alpha_2(\vb{x},t)\J^\dag_{\xi}(t=0)}{\Omega}\nn \\
  &\xrightarrow{t\to\infty}&\sqrt{Z^\alpha_1}\sqrt{Z^\alpha_2}  \sum_{i=0,1}  \psi^{\alpha}_{W_i}(\vb{r})\A_{W_i;\xi} \, e^{-W_{i} t},
  \quad\quad \A_{W_i;\xi}\equiv  \matrixel*{W_i}{\J^\dag_\xi(0)}{\Omega},
\eeqa
 where $W_0$ and $W_1$ are lowest two energies of this coupled channel system.
To extract the $2\times 2$ potential matrix, we need to determine $\A_{W_{0,1};\xi}$ for two linearly independent $J^\dagger_\xi$,
as well as $W_{0,1}$.

The $R$-correlator in the channel $\alpha$, defined by
\beqa
  R^{\alpha}_\xi(\vb{r},t)&\equiv& \frac{F^{\alpha}_\xi(\vb{r},t)}{e^{-m^\alpha_1 t} e^{-m^\alpha_2 t}} \simeq
  \sum_{n,\beta} \A^\alpha_{W_n;\beta,\xi} \psi^{\alpha}_{W_n;\beta }(\vb{r})\, e^{-\Delta^\alpha W_{\xi,n}\,t},
\eeqa
where we take $t>t_\mathrm{threshold}^\mathrm{(inela)}$  in the right-hand side with $\Delta^\alpha W_{\xi,n}\equiv W_{\xi,n}-m_1^\alpha-m_2^\alpha$, satisfies 
\beqa
  \qty(\frac{\nabla^2}{2\mu^\alpha}-\pdv{t}+\frac{1+3{\delta^\alpha}^2}{8\mu^\alpha}\pdv[2]{t})R^{\alpha}_\xi(\vb{r},t) &\simeq& \sum_\beta \tilde{\Delta}^{\alpha\beta}(t)\int\dd[3]{\vb{r'}}U^{\alpha\beta}(\vb{r,r'})R^{\beta}_\xi(\vb{r}',t)
  \label{eq:def_coupled_potential_tdep}
\eeqa
up to $\O((\Delta W)^2)$ as in the single-channel case,  where
\beqa
 \tilde{\Delta}^{\alpha\beta}(t)= \sqrt{\frac{Z^\beta_1 Z^\beta_2}{Z^\alpha_1 Z^\alpha_2}}\frac{\mathrm{e}^{-(m_{1}^{\beta}+m_{2}^{\beta})t}}{\mathrm{e}^{-(m_{1}^{\alpha}+m_{2}^{\alpha})t}},
 \eeqa
 which is needed to correct differences  in masses and $Z$-factors between two channels.
Denoting the left-hand side of eq.(\ref{eq:def_coupled_potential_tdep}) as $\K^\alpha_\xi(\vb{r},t)$, the LO potential is extracted  as
\beqa
 \label{eq:calc_coupled_LOpotential_tdep}
\left(
\begin{array}{cc}
    V^{00}(\vb{r}) & \tilde{\Delta}^{01}(t)V^{01}(\vb{r}) \\
    \tilde{\Delta}^{10}(t)V^{10}(\vb{r}) & V^{11}(\vb{r})
 \end{array}
 \right)
  =
\left(
\begin{array}{cc}
    \K^{0}_{0}(\vb{r},t) & \K^{0}_{1}(\vb{r},t)\\
    \K^{1}_{0}(\vb{r},t) & \K^{1}_{1}(\vb{r},t)
 \end{array}
 \right)   
\left(
\begin{array}{cc} 
    R_{0}^{0}(\vb{r},t) & R_{1}^{0}(\vb{r},t)\\
    R_{0}^{1}(\vb{r},t) & R_{1}^{1}(\vb{r},t)
 \end{array}
 \right)^{-1} .
\eeqa
As before, there is no guarantee that the LO potential is Hermitian.

\section{Lattice QCD setup }
\label{sec:LQCD_setup}

\subsection{Operators}
We are interested in the doubly-bottomed tetra-quark state with quantum numbers $I(J^P) = 0(1^+)$, called $T_{bb}$ hereafter. 
The lowest scattering channel with these quantum numbers is the $\B$ ($\bar{B}\bar{B}^*$) channel with threshold near $10600$ MeV,
while the second one is the $\B^*\equiv$ ($\bar{B}^*\bar{B}^*$) channel with a threshold at $45$ MeV above\cite{ParticleDataGroup:2020ssz}.
Since the threshold of the third channel is too far above to contribute low energy states such as $T_{bb}$, 
we only consider $\B$ and $\B^*$ channels in this paper.

Sink operators for  two mesons at a distance $\vb{r}$ 
with a total spin $S=1$ and a total iso-spin $I=0$ are taken as
\beqa
  \B_j &\equiv& \sum_{\vb{x}}\underbrace{(\bar{u}(\vb{y})\gamma_5 b(\vb{y}))}_{\bar{B}(\vb{y})}\underbrace{(\bar{d}(\vb{x})\gamma_j b(\vb{x}))}_{\bar{B}^*(\vb{x})}-[u\leftrightarrow d],\quad \vb{y}\equiv \vb{x+r},\\
  \B_j^* &\equiv&\epsilon_{jkl}\sum_{\vb{x}}\qty(\bar{u}(\vb{y})\gamma_ k b(\vb{y}))\qty(\bar{d}(\vb{x})\gamma_l b(\vb{x}))-[u\leftrightarrow d] ,
\eeqa
where $j, k, l$ are spatial vector indices. 
At the source, interchanges between $q\leftrightarrow \bar{q}$ are made   for $q=u,d,b$,
together with uses of  wall sources for $q=u,d$.

In addition to these two meson operators,
we introduce an operator made of two diquarks, called $\D$, at the source as
\beqa
 \mathcal{D}_j^{\dagger} \equiv \left(\epsilon^{a b c} \bar{b}^{b}(s_0) \gamma_{j} C \bar{b}^{c}(s_0)\right)\left(\epsilon^{a d e} d^{d}(s_0) C \gamma_{5} u^{e}(s_0)\right)-[u \leftrightarrow d],
\eeqa
where $a, b, c, \dots$ denote color indices, $C = \gamma_4\gamma_2$ is the charge-conjugation matrix, and the argument $s_0$ in the quark field denotes a source point\cite{Leskovec:2019ioa}.
 
A reason for a use of  the diquark at the source is as follow.
If  we perform a coupled-channel analysis with $\B^\dag$ and ${\B^*}^\dag$ source operators,
an inverse matrix in  eq.(\ref{eq:calc_coupled_LOpotential}) or (\ref{eq:calc_coupled_LOpotential_tdep}) becomes singular,
probably  because   $\B^\dag$ and ${\B^*}^\dag$ source operators  couple mainly to the same state,  
as seen Fig.5 of Ref.\cite{Leskovec:2019ioa}.
 To overcome this difficulty, we introduce the diquark-type source $\D^\dag$, which probably couples to a different state.
 We then perform the coupled-channel analysis 
for the $R$-correlators (or the NBS wave functions) with $\B$ and $\B^*$ as sink operators, and $\B^\dag$ and $\D^\dag$ as source operators,
which leads to more stable results than  ${\B^*}^\dag$ and $\D^\dag$ sources.

\subsection{Light Quark Propagators}
In this work, we impose exact isospin symmetry on $u,d$ quarks,  
so that propagators for both quarks are identical.
In our study, we employ the Wilson-Clover operator for the quark, given by
\beqa
  D(x| y) =   \delta_{x, y}&-&\kappa \sum_{\mu}\left\{\left(1-\gamma_{\mu}\right) U_{\mu}(x) \delta_{x+\hat{\mu}, y}+\left(1+\gamma_{\mu}\right) U_{\mu}^{\dagger}(x-\hat{\mu}) \delta_{x-\hat{\mu}, y}\right\}\nn \\
   &-&\kappa\, c_{\mathrm{sw}} \frac{1}{2} \sum_{\mu, \nu} \frac{\left[\gamma_{\mu}, \gamma_{\nu}\right]\left[\Delta_{\mu}, \Delta_{\nu}\right]}{2},
\eeqa  
where $\Delta_\mu$ in the clover term are symmetric covariant difference operator, defined by
\beqa
\Delta_\mu f (x) = U_{\mu} (x) f(x+\hat\mu) - U_{\mu}^\dagger (x-\hat\mu) f(x-\hat\mu), 
\eeqa
and $\hat\mu$ is a unit vector in the $\mu$ direction with a length $a$, where $a$ is a lattice spacing.
See  Sec.~\ref{sec:conf} for parameters $\kappa,c_\mathrm{sw}$ used in this study.
As mentioned before, we use wall sources for light quarks. 

\subsection{Heavy Quark Propagator}
As long as the relativistic lattice fermion is used, 
$am_Q \ll 1$ is required to keep lattice artifact small, where $m_Q$ is a quark mass.
This condition, however,  is badly violated for the $b$ quark in our simulations, since $m_b\approx4.2$ GeV and $a\approx0.09$ fm ($1/a\simeq 2$ GeV).
Therefore, we cannot treat the $b$ quark relativistically on a lattice. 
Fortunately, since the typical velocity of the $b$ quark inside a hadron is $v^2\sim 0.1$\cite{Thacker:1990bm}, and thus sufficiently non-relativistic, 
we can treat the $b$ quark in the Non-Relativistic QCD (NRQCD) approximation.
The NRQCD approximation improves the static approximation,
by including effects of moving $b$ quarks in space,
which seem to give a non-negligible contribution to the binding energy of the
tetra-quark state\cite{Bicudo:2015kna,Leskovec:2019ioa}.  

In the NRQCD, we evaluate a time-evolution of the heavy quark propagator according to non-relativistic dynamics
using a Hamiltonian without $b$ quark mass term. The NRQCD Hamiltonian at the tree level is obtained from the QCD Hamiltonian by the Foldy-Wouthuysen-Tani (FWT) transformation\cite{Foldy:1949wa,taniConnectionParticleModels1951} designed to be block-diagonal up to $\O(v^n)$ in spinor space as
\beqa
  \H_{\rm QCD} \to \R\H_{\rm NRQCD}\R^\dag
  \simeq \R
\left(  \begin{array}{cc}
  \H_\psi & 0 \\ 0 & \H_{\chi^\dag}
 \end{array} \right)
 \R^\dag,
\eeqa
where $\R$ is the FWT transformation matrix. The propagator for the particle field $\psi$ moving in the positive direction can be approximated as
\beqa
  \label{eq:NRQCDpropag}
  D^{-1}(x|y) \to \R\contraction{}{\psi}{(x)}{\psi}\psi(x)\psi^\dag(y)\R^\dag
  \simeq \R
\left(  \begin{array}{cc}  
G_\psi(x|y) & 0 \\ 0 & 0
 \end{array} \right)
\R^\dag\,\theta(x_4-y_4),
\eeqa
and the two-spinor NRQCD propagator $G_\psi$ is evolved in time by $\H_\psi\equiv \H_0+\delta \H$ on a lattice as\cite{Lepage:1992tx}
\beqa
  \label{eq:NRQCDpropag_time_evol}
  G(\vb{x}, t+1|s_0)
  &=& \left(1-\frac{ \H_{0}}{2 n}\right)^{n}\left(1-\frac{ \delta \H}{2}\right) U_{4}^{\dagger}(x)\left(1-\frac{\delta \H}{2}\right)\left(1-\frac{\H_{0}}{2 n}\right)^{n} G(\vb{x}, t|s_0) \nn \\
  &+&s_0(\vb{x})\delta_{t+1,0},
\eeqa
where $s_0$ is a source vector defined previously, and $n=2$ is a stabilization parameter for numerical calculations. This calculation requires much smaller computational costs than solving linear equations for relativistic quark propagators.
In this work, we use the block-diagonal Hamiltonian up to $\O(v^4)$ \cite{Ishikawa:1997sx}, given on a lattice as 
\beqa
    \H_\psi&=& \H_0+\sum_i c_i\delta \H^{(i)}, \quad
      \H_0  = -\frac{1}{2M}\Delta^{(2)}, \quad\nn \\
      \delta \H^{(1)} &=& -\frac{1}{2 M} {\bf \sigma} \cdot \vb{B},  \quad
      \delta \H^{(2)} = \frac{i}{8 M^{2}}(\Delta \cdot \vb{E}-\vb{E} \cdot \Delta),  \quad
      \delta \H^{(3)} = -\frac{1}{8 M} {\bf\sigma} \cdot(\Delta \times \vb{E}-\vb{E} \times \Delta),\nn \\
      \delta \H^{(4)} &=& -\frac{1}{8 M^{3}}(\Delta^{(2)})^{2} ,\quad
      \delta \H^{(5)} =\frac{1}{24 M}\Delta^{(4)} ,\quad 
      \delta \H^{(6)} =-\frac{1}{16 n M^{2}}(\Delta^{(2)})^{2},~~~
\eeqa
where $M$ is the bare heavy-particle mass,  $c_i=1$ at the tree level in perturbation theory,
$\Delta,\Delta^{(2)},\dots$ are discretized symmetric covariant derivatives in space, and the chromo-electromagnetic field $\vb{E,B}$ are given by the standard clover-leaf definitions. 
The FWT transformation matrix is also given up to $\O(v^4)$ \cite{Ishikawa:1997sx} as
\beqa
    \R &=&1+\sum_{i} \R^{(i)},\quad\quad\nn \\
      \R^{(1)} &=& -\frac{1}{2 M} \gamma \cdot {\Delta}, \quad
      \R^{(2)} = \frac{1}{8 M^{2}}\Delta^{(2)}, \quad
      \R^{(3)} = \frac{1}{8 M^{2}}\bm{\Sigma} \cdot \vb{B}, \quad
      \R^{(4)} = -\frac{i}{4 M^{2}}\gamma_{4} \bm{\gamma} \cdot \vb{E}.~~
\eeqa

In our study, all link variables are rescaled as $U_\mu\to U_\mu/u_0$, 
in order to include perturbative corrections by the tadpole improvement\cite{Lepage:1992tx},
where $u_0$ is determined from an average of the plaquette $U_P$ as 
\beqa
  \label{eq:link_expval}
  u_0=\left\{ \frac{1}{3} \Tr U_P \right\}^{1/4}.
\eeqa

In the lattice NRQCD, the ground state energy obtained from a behavior of the two-point function in time represents 
the interaction energy, not the hadron mass itself,  since the quark mass term is removed from the NRQCD Hamiltonian. 
Therefore, a correlation function with non-zero momentum behaves at large $t$ as
\beqa
\left\langle  H_X(\vb{p},t)H^\dag_X(\vb{p},0)\right\rangle \xrightarrow{t\to\infty} e^{-E_X({\vb{p}})t},\quad H_X(\vb{p})\equiv \sum_{\vb{x}}H_X(\vb{x}) e^{-i\vb{px}},
\eeqa
where $E_X(\vb{p})=\sqrt{\vb{p}^2+(M_X^\mathrm{kin})^2}-\delta$ with the ${\bf p}$ independent energy shift $\delta$.
Since  this energy shift $\delta$, equal to the bare quark mass at the tree level, usually suffers from
large perturbative corrections, 
 we directly estimate a (kinetic) mass of the hadron $X$ without determining $\delta$ as 
\beqa
  M_X^{\mathrm{kin}}= \frac{\vb{p}^2-(E_X(\vb{p})-E_X(\vb{0}))^2}{2(E_X(\vb{p})-E_X(\vb{0}))}.
  \label{eq:Mkin}
\eeqa

\subsection{Configurations\label{sec:conf}}
We have employed the $(2+1)$ flavor full QCD configurations, generated by the PACS-CS Collaboration\cite{PACS-CS:2008bkb} with the Iwasaki gauge action and the Wilson-Clover light quark action at $a\approx0.09$ fm.
For the wall source, gauge configurations are fixed to the Coulomb gauge. 
We estimate statistical errors by the jackknife method, with a bin size 20,  using 400 configurations on each quark mass.
Parameters for gauge ensembles and hadron masses measured in this work are listed in tables \ref{tab:conf_parameters} and \ref{tab:conf_physobs},
respectively.

\begin{table}[hbtp]
  \centering
  \begin{tabular}{c|cccccccc}
    \hline
    Configuration & $V_\mathrm{lat}=L_s^3\times L_t$ & $a$ [fm]  & $L_s$ [fm] & $\kappa_{ud}$ & $\kappa_s$ & $c_{\rm sw}$ & $M_b$ & $u_0$ \\
    \hline \hline
      PACS-CS-A  & $32^3\times 64$ & $0.0907(13)$ & $2.902(42)$ & $0.13700$ & $0.13640$ & $1.715$ & $1.919$ & $0.868558(42)$ \\
      PACS-CS-B  & $32^3\times 64$ & $0.0907(13)$ & $2.902(42)$ & $0.13727$ & $0.13640$ & $1.715$ & $1.919$ & $0.868793(43)$ \\
      PACS-CS-C & $32^3\times 64$ & $0.0907(13)$ & $2.902(42)$ & $0.13754$ & $0.13640$ & $1.715$ & $1.919$ & $0.869005(44)$ \\
    \hline
  \end{tabular}
  \caption{Parameters for gauge ensembles. The bare $b$ quark mass $M_b$ is taken to satisfy $M_{b\bar{b}}^{\rm spinavg}\approx 9450$ {MeV} within errors. The expectation value of the link variable $u_0$ defined in eq.(\ref{eq:link_expval}) is used for the tadpole improvements.\label{tab:conf_parameters}}
\end{table}

\begin{table}[hbtp]
  \centering
  \begin{tabular}{c|cccc}
    \hline
    Configuration & $m_\pi$ [MeV] & $m_\rho$ [MeV] & $M_{\bar{B}}^{\rm spinavg}$ [MeV] & $\Delta E_{\bar{B}\bar{B}^*}$ [MeV] \\
    \hline \hline
      PACS-CS-A & $701(1)$ & $1102(1)$ & $5440(174)$ & $49.4(2.6)$ \\
      PACS-CS-B & $571(0)$ & $1011(1)$ & $5382(269)$ & $44.9(1.6)$ \\
      PACS-CS-C & $416(1)$ & $920(3)$ & $5332(220)$ & $42.7(3.9)$ \\
    \hline
  \end{tabular}
  \caption{Hadron masses measured on each ensemble. The $B$ meson mass $M_{\bar{B}}$ is determined by the kinetic mass, and the spin-averaged is made as $\frac{1}{4}M_{\bar{B}}+\frac{3}{4}M_{\bar{B}^*}$. The energy splitting $\Delta E_{\bar{B}\bar{B}^*}$ is defined by $\Delta E_{\bar{B}\bar{B}^*}\equiv E_{\bar{B}^*}(\vb{0})-E_{\bar{B}}(\vb{0})$.\label{tab:conf_physobs}}
\end{table}

Comments on measured hadron masses are in order. 
\begin{itemize}
  \item   While an individual mass of $\bar{B}$ or $\bar{B}^*$ has a sizable statistical error due to a use of data at non-zero ${\bf p}$ in eq.~\eqref{eq:Mkin}, we can determine the mass splitting between them from $E_{\bar{B}^*}(\vb{0})-E_{\bar{B}}(\vb{0})$,
  which does not require noisy data at  non-zero ${\bf p}$.
  In the table, we also list the spin average mass $M_{\bar B}^{\rm spinavg}\equiv \frac{1}{4}M_{\bar{B}}+\frac{3}{4}M_{\bar{B}^*}$. 
  For calculations of potentials, we need to use $M_{\bar B}$ and $M_{\bar{B}^*}$ separately.
  
  \item  Values of $\bar B$ meson mass in the table are consistent with an experimental value  $M_{\bar{B}}^\mathrm{spinavg}=5313$ {MeV} \cite{ParticleDataGroup:2020ssz} within large statistical errors at 3 light quark masses, and we expect  that this agreement holds even at the physical pion mass.
 Thanks to smaller statistical errors, on the other hand,
 we observe a tendency that the mass splitting $\Delta E_{\bar{B}\bar{B}^*}$  decreases as the pion mass decreases and it becomes smaller than 
 an experimental value $\Delta E_{\bar{B}\bar{B}^*}=45$ {MeV}\cite{ParticleDataGroup:2020ssz} at the physical pion mass.
 Among  possible reasons for this, 
 it is most likely that $c_1=1$ with the tadpole improvement is not good enough as a coefficient of $\delta\H^{(1)}$ in the NRQCD Hamiltonian,
 which is the LO term in the NRQCD power counting responsible for the  spin splitting. 
 Therefore we expect $10$-$20\%$ systematic errors for the spin splittings
 at the tree level coefficient even with the tadpole improvement.
\end{itemize}

In this work, scattering quantities are calculated on 3 different pion masses, and then extrapolated to the physical point defined by $m_\pi\approx140$ {MeV}.

\section{Numerical results}
\label{sec:numerical_result}

\subsection{Leading-Order Potential}
\subsubsection{Single-channel case}
\begin{figure}[tbh]
\centering
\vskip -8cm
\includegraphics[width=1.0\linewidth,angle=0]{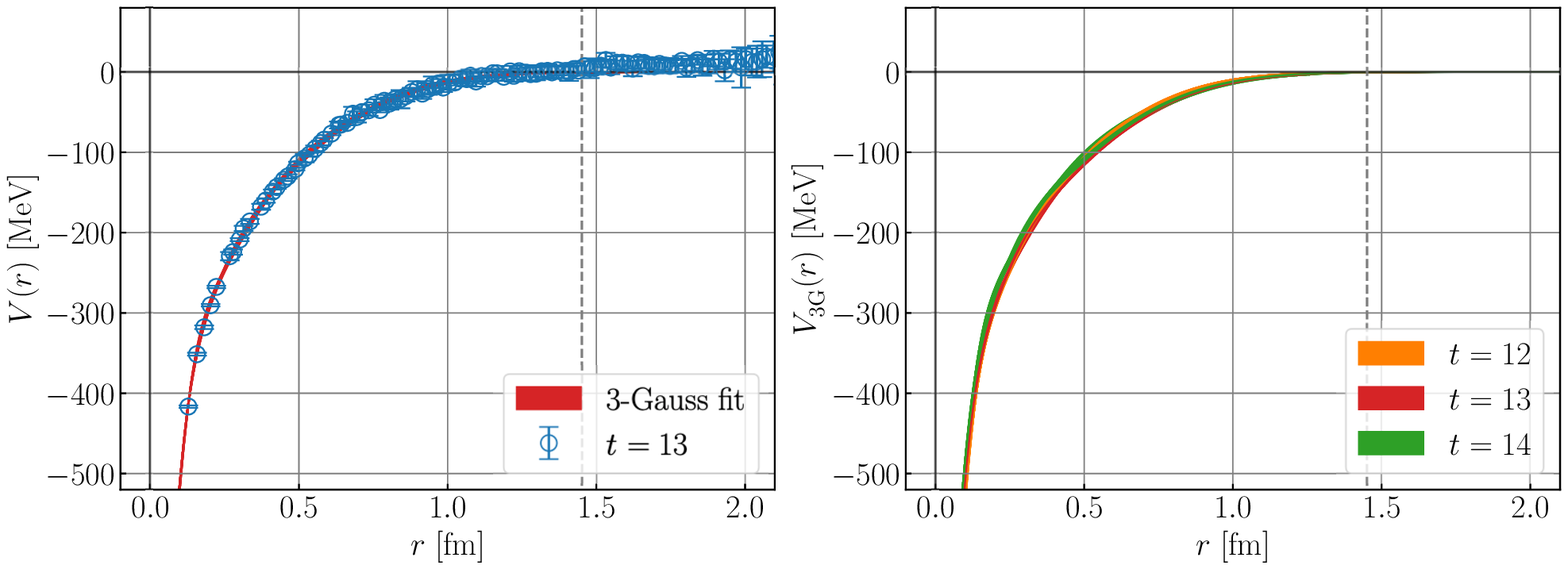}
\vskip -8cm
\caption{(Left)  A lattice result of the potential at $t=13$ (blue circles), together with the 3-Gauss fit by a red line. (Right)   3-Gauss fits at $t=12,13,14$.  A gray dashed-line indicates $r=L/2$. \label{fig:potential_fit_single}}
\end{figure}

In this subsection, assuming that the $T_{bb}$  couples only to the $\B$ channel, we compute the $S$-wave\footnote{The NBS wave function is projected to the $A_1^+$ representation of the cubic group, where 
we ignore higher partial waves such as $\ell=4,6,\cdots $.} LO potential according to eq.(\ref{eq:LO_potential_tdep}). 
Fig~\ref{fig:potential_fit_single} (Left) shows the one at $m_\pi\simeq 700$ MeV (PACS-CS-A) and $t=13$.
The potential between $\bar B$ and $\bar B^*$ mesons is attractive at all distances and it becomes zero within errors at 
distances larger than 1.0 fm,  which is smaller than $L_s/2\simeq 1.45$ fm.  Thus the interaction is sufficiently short-ranged to be confined within the box, so that finite size effect to the potential is expected to be small. 
To fit data of the potential, we use a 3-Gauss function given by
\beqa
  \label{eq:3gauss}
  V_\mathrm{3G}(r)&=& \V_0 e^{-r^2/\rho_0^2}+\V_1 e^{-r^2/\rho_1^2}+\V_2 e^{-r^2/\rho_2^2},
\eeqa
where $\V_i$ and $\rho_i$ are fit parameters. We show fit results to lattice data at $t=12$--$14$ in Fig.~\ref{fig:potential_fit_single} (Right),
whose time-dependence is negligibly small, indicating that contaminations from inelastic states are well under control. 
Thus we have employed the potential at $t=13$ for our main analysis, whose fit parameters are given in Tab~\ref{tab:3G_fitparams_single}. 

\begin{table}[h]
\centering
\begin{tabular}{ccc}
\hline
$V^{}$ & $\V_i$ [MeV] & $\rho_i$ [fm]\\
\hline
$i=0$& $-482(11)$& $0.088(0.002)$\\
$i=1$& $-185(8)$& $0.218(0.004)$\\
$i=2$& $-236(2)$& $0.583(0.002)$\\
\hline
\end{tabular}
\caption{3-Gauss fit parameters at $t=13$.\label{tab:3G_fitparams_single}}
\end{table}

\subsubsection{Coupled-channel case}
\begin{figure}[htb]\centering
\vskip -5cm
 \includegraphics[width=0.65\linewidth,angle=0]{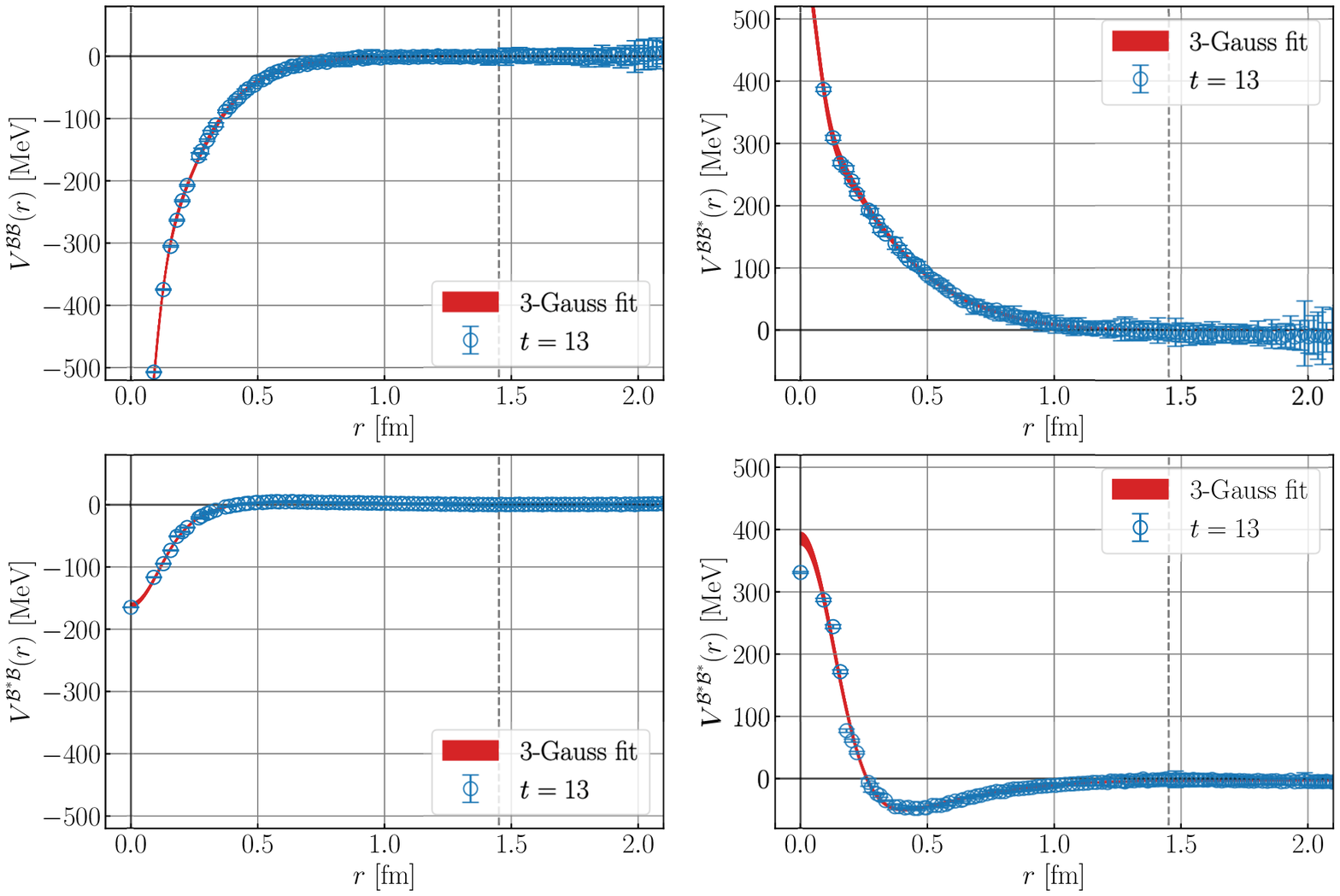}
 \vskip -7cm
 \includegraphics[width=0.65\linewidth,angle=0]{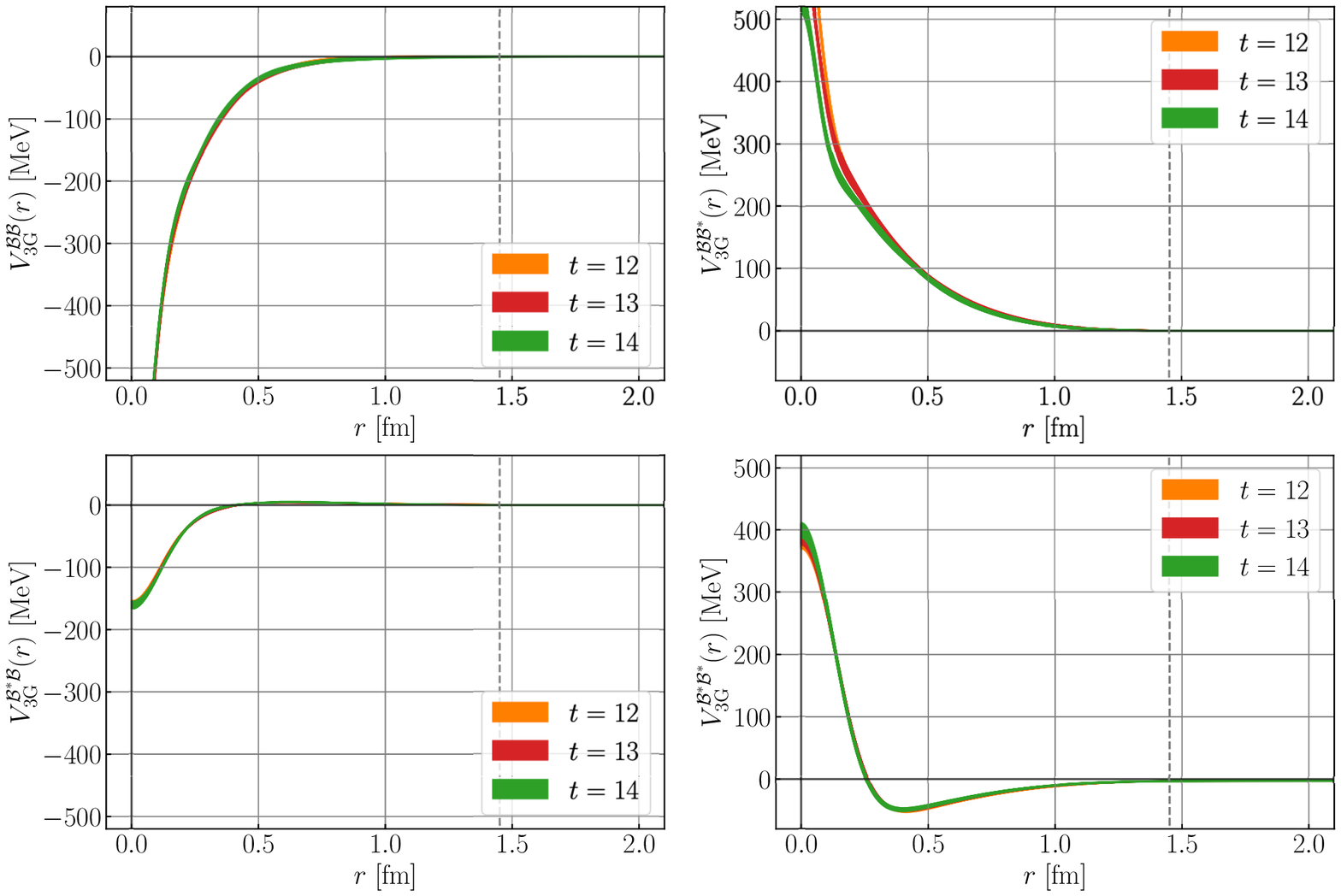}
 \vskip -4cm
 \caption{(Upper) $2\times 2$ coupled-channel potentials (blue circles) at $t=13$, together with 3-Gauss fits by red lines. 
  (Lower) 3-Gauss fits at $t=12,13,14$. 
   \label{fig:potential_fit_coupled}}
\end{figure}

 We now consider a case that the $T_{bb}$ couples to $\B$ and $\B^*$ channels.
 In this situation, we compute the $S$-wave LO potential using eq.(\ref{eq:calc_coupled_LOpotential_tdep}). 
 Fig.~\ref{fig:potential_fit_coupled} (Upper) show $2\times 2$ coupled-channel potentials at $m_\pi\simeq 700$ MeV (PACS-CS-A ) and $t=13$,
 which become zero within errors at $r\gtrapprox 1.0$ fm, together with 3-Gauss fit by red lines.
 As before, we thus confirm that  interactions in this channel are sufficiently short-range, so that possible finite size effects are expected to be small.

A diagonal potential,  $V^{\B\B}$,  is attractive at distances smaller than 0.8 fm, while another one, $V^{\B^*\B^*}$,   has a repulsive core at short distances surrounded by an attractive pocket  at $r\simeq 0.4$ fm. 
On the other hand,
magnitudes of off-diagonal interactions between $\B$ and $\B^*$ channels are comparable to those of diagonal interactions,
showing that a channel coupling between $\B$ and $\B^*$ is significant. 
This observation suggests an importance of a coupled-channel analysis or  conversely
a possibility that a single-channel analysis may contain large systematic uncertainties. 
In addition, we have observed that  Hermiticity  of  the $2\times 2$ potential matrix is badly broken:
two off-diagonal components are very different.
We speculate that the leading order approximation for the original non-local coupled-channel potential, which should be Hermitian,  causes 
this large violation of Hermiticity, suggesting strong non-locality of the coupled-channel potential in this system,
which is consistent with our observation that off-diagonal interactions are significant.

Since the standard scattering analysis requires the unitarity of the $S$-matrix, which is guaranteed by Hermitian potentials, 
we can not perform the coupled channel analysis for scatterings above  the $\B^*$ threshold. 
In this paper, however, we still employ coupled-channel potentials for a scattering analysis in the $\B$ channel below the $\B^*$ threshold,
in order to partly incorporate non-locality caused by  {\it off-shell} $\B^*$ propagations.
Details of such an analysis will be given in Sec.~\ref{sec:scattering_analysis}. 

Fig.~\ref{fig:potential_fit_coupled} (Lower) presents 3-Gaussian fits to lattice data at  $t=12$--$14$. 
An off-diagonal component $V^{\B\B^*}$ show a detectable time-dependence at the short distance,
which however is found to give tiny effects on scattering quantities.
We therefore concluded that contributions from  inelastic states are well under control, and 
we employ  $t=13$ data in our main analysis. 
Tab.\ref{tab:3G_fitparams_coupled} gives fit parameters of the coupled channel potential at $t=13$.

\begin{table}[h]
\centering
\begin{tabular}{ccc}
\hline
$V^{\B\B}$& $\V_i$ [MeV]& $\rho_i$ [fm]\\
\hline
$i=0$& $-491(60)$& $0.092(0.013)$\\
$i=1$& $-254(64)$& $0.255(0.095)$\\
$i=2$& $-110(144)$& $0.476(0.180)$\\
\hline

\end{tabular}\quad\quad
\begin{tabular}{ccc}
\hline
$V^{\B\B^*}$& $\V_i$ [MeV]& $\rho_i$ [fm]
\\
\hline
$i=0$& $302(38)$& $0.086(0.007)$\\
$i=1$& $138(47)$& $0.289(0.090)$\\
$i=2$& $170(65)$& $0.578(0.063)$\\
\hline
\end{tabular}\quad\quad
\\

\begin{tabular}{ccc}
\hline
$V^{\B^*\B}$& $\V_i$ [MeV]& $\rho_i$ [fm]\\
\hline
$i=0$& $-109(17)$& $0.147(0.125)$\\
$i=1$& $-61.1(16.2)$& $0.288(0.051)$\\
$i=2$& $-9.27(4.95)$& $0.820(0.188)$\\
\hline
\end{tabular}\quad\quad
\begin{tabular}{ccc}
\hline
$V^{\B^*\B^*}$& $\V_i$ [MeV]& $\rho_i$ [fm]\\
\hline
$i=0$& $456(26)$& $0.181(0.005)$\\
$i=1$& $-76.2(6.1)$& $0.657(0.043)$\\
$i=2$& $-1.53(1.45)$& $1.385(0.099)$\\
\hline
\end{tabular}
\caption{3-Gauss fit parameters at $t=13$.
\label{tab:3G_fitparams_coupled}}
\end{table}

\subsubsection{Pion mass dependence}
\begin{figure}[hbt]
\centering
\vskip -6cm
\includegraphics[width=\linewidth]{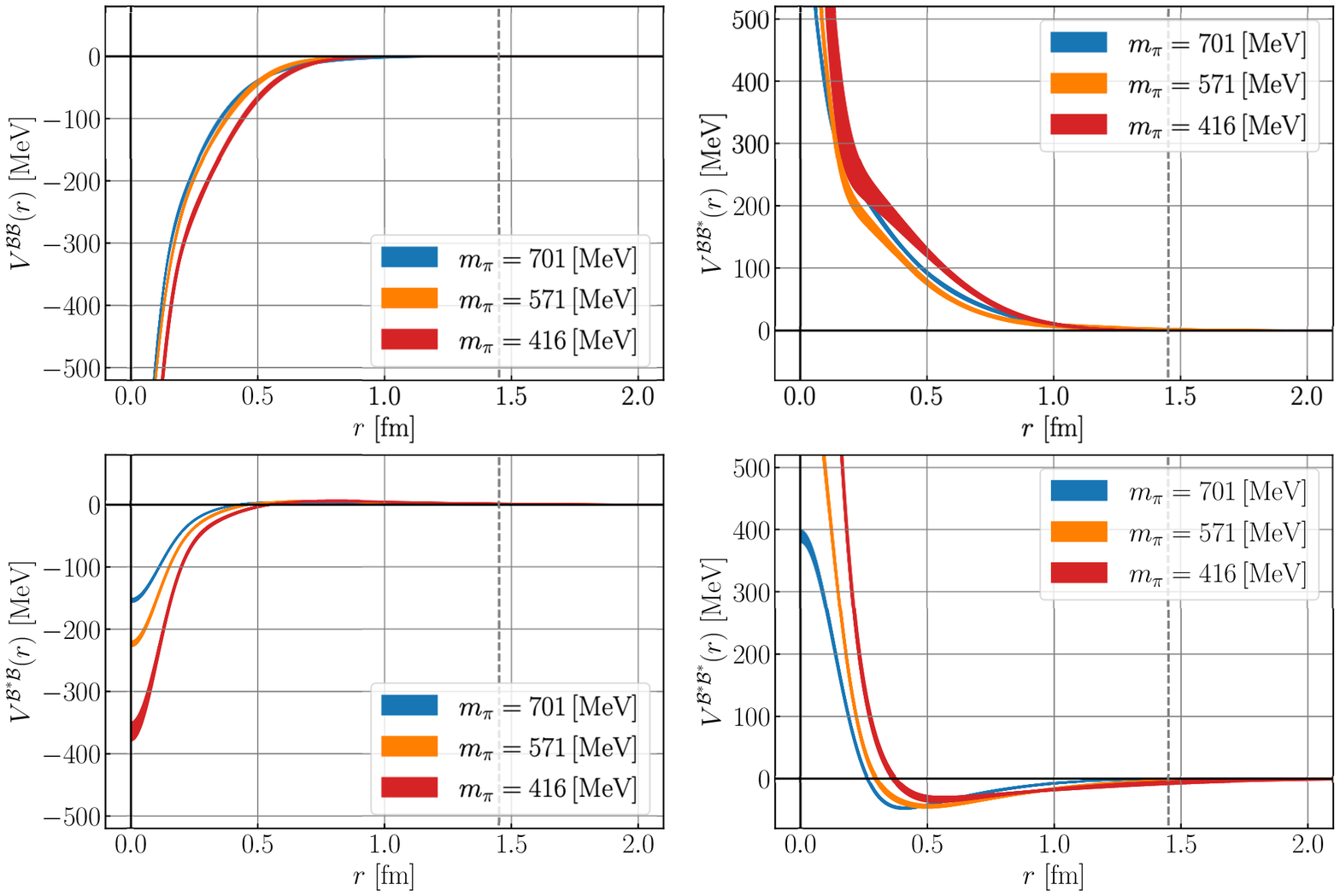}
\vskip -6cm
\caption{Fit results $V^{\alpha\beta}_{\rm 3G}(r)$ at $m_\pi=701$ MeV (blue), 571 MeV (orange) and 416 MeV (red) at $t=13$.\label{fig:coupled_potential_mpiall}}
\end{figure}

Fig.~\ref{fig:coupled_potential_mpiall} compares potentials at three different pion masses,  $m_\pi=701,571,416$ MeV.
As the pion mass gets smaller,  both diagonal and off-diagonal potentials become stronger and more long-ranged. 
This suggests that a mixing effect between $\B$ and $\B^*$ increases toward the physical pion mass, so that
the coupled channel analysis may be mandatory even below the $\B^*$ threshold.
In this study, physical observables such as scattering phase shifts and a binding energy, not potentials themselves,  are extrapolated from results at three heavier pion masses to the physical pion mass, $m_\pi = 140$ MeV. 

\subsection{Scattering Analysis\label{sec:scattering_analysis}}

\subsubsection{Inclusion of virtual $\B^*$ effects}
In the following analysis, we restrict ourself to scatterings only in the $\B$ channel below the $\B^*$ threshold, where
the $\B^*$ channel virtually appear as intermediate states, even though we employ the $2\times 2$ coupled channel potential matrix in the analysis. 
This kind of situation has been analyzed in \cite{Aoki:2011gt}, which shows that effects of virtual $\B^*$ states appear as non-locality of the effective potential in the $\B$ channel.
Explicitly, the  coupled channel Schr\"odinger equation between $\B$ and $\B^*$ becomes an effective single channel 
Schr\"odinger equation in the $\B$ channel as
\beqa
(H_0 + U_{{\rm eff},E}^{\B\B}) \Psi_B = E \Psi_B,
\eeqa 
where
\beqa
  U_{{\rm eff},E}^{\B\B}(\vb{x},\vb{y})&=& V^{\B\B}(\vb{x})\delta(\vb{x-y})+ V^{\B\B^*}(\vb{x})G_E^{\B^*\B^*}(\vb{x,y})V^{\B^*\B}(\vb{y}),
\eeqa
where $G_E^{\alpha\alpha}(\vb{x,y})=(E-H^{\alpha}_0-V^{\alpha\alpha})^{-1}(\vb{x,y})$ is the full Green function for the energy $E$ in the $\alpha$ channel, and thus the effective potential $U^{\B\B}_{{\rm eff},E}(\vb{x,y})$ explicitly depends on the energy $E$.
In this expression, it is clear that effects of intermediate $\B^*$ states leads to non-locality for $U^{\B\B}_{{\rm eff},E}(\vb{x,y})$ in the second term.
While the original $U(\vb{x,y})$ is defined in QCD,  $U^{\B\B}_{{\rm eff},E}(\vb{x,y})$ contains only a part of non-locality caused by
such intermediate $\B^*$ states with {\it local} interactions $V^{\B^*\B^*}$, $V^{\B\B^*}$ and $V^{\B^*\B}$ for a given energy $E$.
A remaining non-locality comes not only from non-locality of coupled channel potentials but also from virtual channels other than $\B$ and $\B^*$,
latter of which have negligible effects on the scattering in the $\B$ channel below the $\B^*$ threshold, since thresholds of other channels 
are far above from it.

Note that, even though  $U^{\B\B}_{{\rm eff},E}(\vb{x,y})$ is still non-hermitian,  
we can extract real scattering phase shifts in the $\B$ channel  and thus an unitary $S$-matrix, as long as we take {\it real} $E$ below the $\B^*$ threshold\cite{Aoki:2019gqt},
as will be explicitly shown later. 
For the analysis, we employ the coupled channel Lippmann-Schwinger equation to incorporate effects of virtual $\B^*$ to the scattering in the $\B$ channel, which indeed lead to sizable corrections to results obtained in the single channel analysis without virtual $\B^*$ states. 

\subsubsection{Matrix inversion method for the Lippmann-Schwinger equation}
 The Lippmann-Schwinger(LS) equation for the $T$-matrix with the potential matrix reads 
\beqa
  \label{eq:Lippmann_for_Tmat}
  T^{\alpha\beta}(\vb{p}^\alpha_W,\vb{p}_W^\beta) &=& V^{\alpha\beta}(\vb{p}^\alpha_W,\vb{p}_W^\beta) \nn \\
  &+&\sum_\gamma\int\dd[3]{\vb{k}}V^{\alpha\gamma}(\vb{p}^\alpha_W,\vb{k})\frac{1}{(W-E_\mathrm{th}^\gamma)-\vb{k}^2/2\mu^\gamma+i\varepsilon}T^{\gamma\beta}(\vb{k},\vb{p}^\beta_W),
\eeqa
where $p_W^\alpha=\sqrt{2\mu^\alpha(W-E_{\rm th}^\alpha)}$ is the momentum calculated from the total energy $W$ and 
$E_{\rm th}^\alpha=M^\alpha_1+M^\alpha_2$ is a threshold energy for a channel $\alpha$. 

We have employed the matrix inversion method\cite{Haftel:1970zz} to solve the LS equation, 
approximating the momentum integral by a finite sum over Gaussian quadrature points. 
For the $S$-wave component in the partial wave expansion, the LS equation is reduced to
\beqa
  t_{\ell=0}^{\alpha \beta}(k_{i}^{\alpha}, k_{j}^{\beta})= \hat{V}_{\ell=0}^{\alpha \beta}(k_{i}^{\alpha}, k_{j}^{\beta})-\sum_{\gamma, \delta} \sum_{m, n=0}^{N} \hat{V}_{\ell=0}^{\alpha \gamma}(k_{i}^{\alpha}, k_{m})\, \hat{G}_{0}^{\gamma \delta}(k_{m}, k_{n})\, t_{\ell=0}^{\delta \beta}(k_{n}, k_{j}^{\beta}),
\eeqa
where $k_s$ with $s =0,\cdots, N-1$ represents a momentum at the Gaussian quadrature point, while the on-shell momentum $p_W^\alpha$ is stored in $k_N^\alpha$. 
A matrix element of the potential for
a Gaussian expansion of the potential, $V(r)=\sum_k\V_k e^{-r^2/\rho_k^2}$,  is defined as
\beqa
  \hat{V}_0^{\alpha\beta}(k^{\alpha}_i,k^{\beta}_j)\equiv \frac{1}{\sqrt{4\pi}}\sqrt{\frac{\mu^\alpha\mu^\beta}{k^\alpha_i k^\beta_j}}\sum_k \V_k \rho_k\exp[-\frac{1}{4}\rho_i^2(k^\alpha_i+k^\beta_j)^2]\qty(\exp[\rho_k^2 k^\alpha_i k^\beta_j]-1),
\eeqa
while the Green function is given by
\beqa
  \hat{G}^{\gamma \delta}_0(k_{m}, k_{n}) \equiv \delta^{\delta \gamma} \delta_{m n} \times 
  \begin{cases}
    \displaystyle \tilde{w}_{m} \frac{2 k_{m}}{k_{m}^{2}-2 \mu^{\gamma}(W-E_{\mathrm{th}}^\gamma)} & (m=0, \ldots, N-1) \\ 
    \\
 \displaystyle     -\sum_{l=0}^{N-1} \tilde{w}_{l} \frac{2 k_{N}^{\gamma}}{k_{l}^{2}-2 \mu^{\gamma}(W-E_{\mathrm{th}}^\gamma)}+\mathrm{i} \pi & (m=N)
  \end{cases},
\eeqa
where
\beqa
    k_{j} &=& p_{\mathrm {cut }} \tan \left[\frac{\pi}{4}\left(x_{j}+1\right)\right], \
    \tilde{w}_{j} = p_{\mathrm{cut}} \frac{\pi}{4} \frac{w_{j}}{\cos ^{2}\left[\frac{\pi}{4}\left(x_{j}+1\right)\right]}, 
    \quad x_j \in [-1,1]
   \eeqa
with the weight $w_j=\frac{2}{(1-x_{j}^{2})[P_{N}^{\prime}(x_{j})]^{2}}$ for the Gauss-Legendre quadrature used in our calculations.
We have confirmed that physical observables are insensitive to our choice, $N=50$ and $p_{\rm cut}=100$ MeV.
(Results are unchanged within errors for $p_{\rm cut}=1000$ MeV or $N=60$.)
Then, the $T$-matrix is  approximately obtained by a matrix inversion as $t_0=({\bf 1}-VG_0)^{-1} V$, 
where $t_0^{\alpha\beta}(k_N^\alpha,k_N^\beta)$ corresponds to the on-shell $T$-matrix $t_0^{\alpha\beta}(W)$.

\subsubsection{$T$-matrix and bound states}
The scattering phase shift can be extracted from the on-shell $T$-matrix as 
\beqa
  \frac{t_0^{\B\B}(W)}{p_W^{\B}}= \frac{-1}{\pi}\frac{1}{p^\B_W\cot\delta^{\B\B}_0(W)- i p_W^{\B}},
\eeqa
and then $p\cot\delta$ is parametrized by the Effective Range Expansion(ERE) as
 \beqa
  \label{eq:ERE_pcotdelta}
  p^\B_W\cot\delta^{\B\B}_{0}(W)= -\frac{1}{a_0}+\frac{r_{\rm eff,0}}{2}(p^\B_W)^2+\O\left((p^\B_W)^4\right)\,.
\eeqa
where $a_0$ is the scattering length and $r_{\rm eff,0}$ is the effective range.

Since bound states correspond to poles of the $T$-matrix in a negative $(p^\B_W)^2 $ axis,
we have to solve the LS equation at $(p^\B_W)^2< 0$  in order to find such poles. 
Alternatively, using \eqref{eq:ERE_pcotdelta},
we may search an intersection between  the ERE $p\cot\delta$ and a bound state condition $-\sqrt{-p^2}$
at $(p^\B_W)^2<0$, which gives a pole at $p=+ i p_{\rm BS}^{}$ in the upper half complex $p^\B_W$ plane.
In addition, for a pole of a physical bound state,  $p\cot\delta$ must cross $-\sqrt{-p^2}$ from below as \cite{Iritani:2017rlk}
\beqa
  \label{eq:bound_condition}
  \left.\dv{p^2}\left[p \cot \delta(p)-(-\sqrt{-p^{2}})\right]\right|_{p^{2}=-p_{\rm BS}^{2}}\ < \ 0\,.
\eeqa

\subsubsection{Results}
\begin{figure}[bth]
\centering
\vskip -4cm
\includegraphics[width=0.8\linewidth,angle=0]{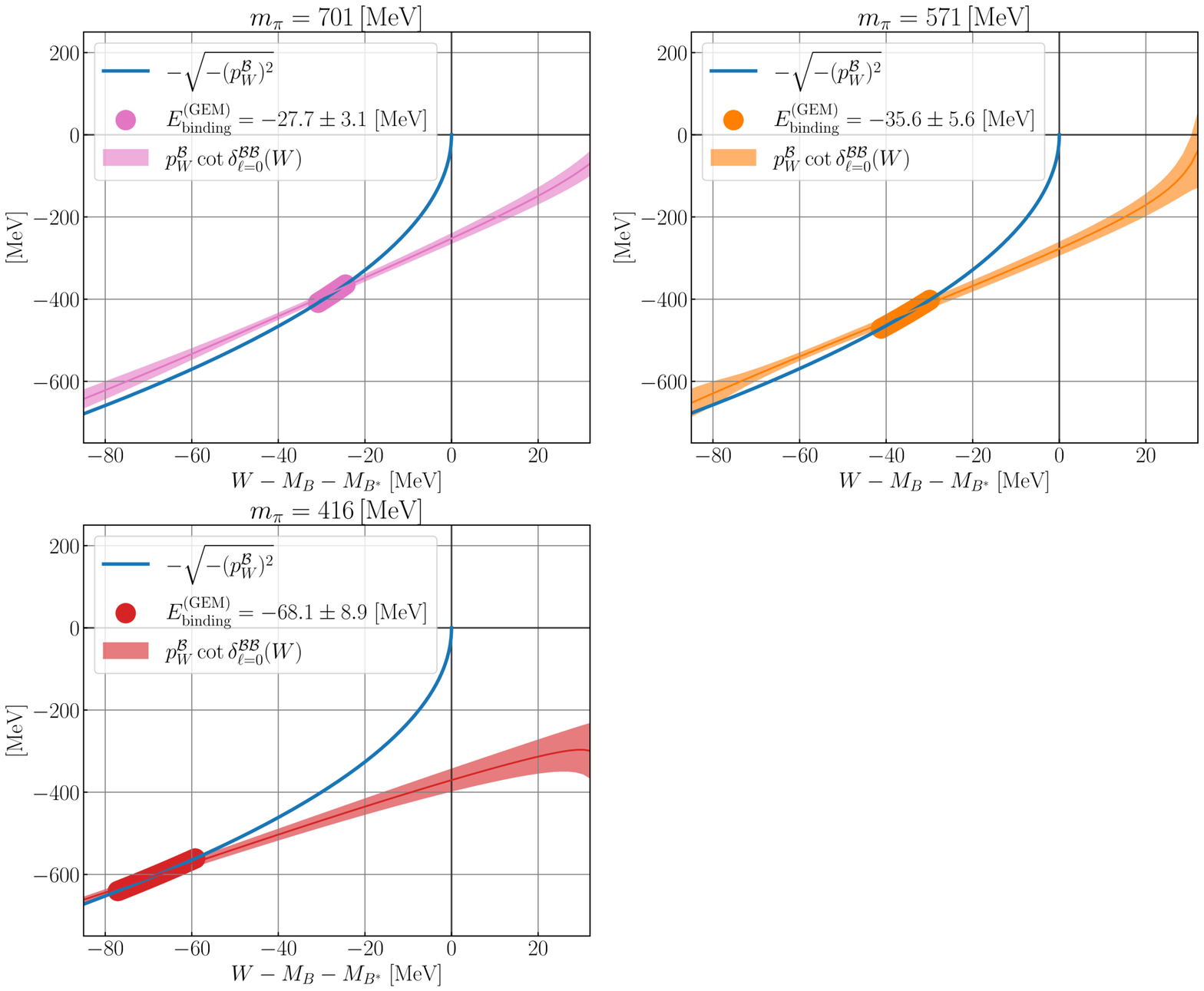}
\vskip -4cm
\caption{Results of  $p\cot\delta(W)$ from the LS equations as a function of  $W <E_{\rm threshold}^{\B^*}$,
together  with $-\sqrt{-p^2}$ by the blue solid line.
A thick line along the $-\sqrt{-p^2}$ curve represents the binding energy calculated by the GEM, which agrees well  with the intersection
corresponding to a pole of the $T$-matrix.\label{fig:pacscs_LIPPvsGEM}}\end{figure}

\begin{figure}[htb]
\centering
\vskip -3cm
\includegraphics[width=0.52\linewidth]{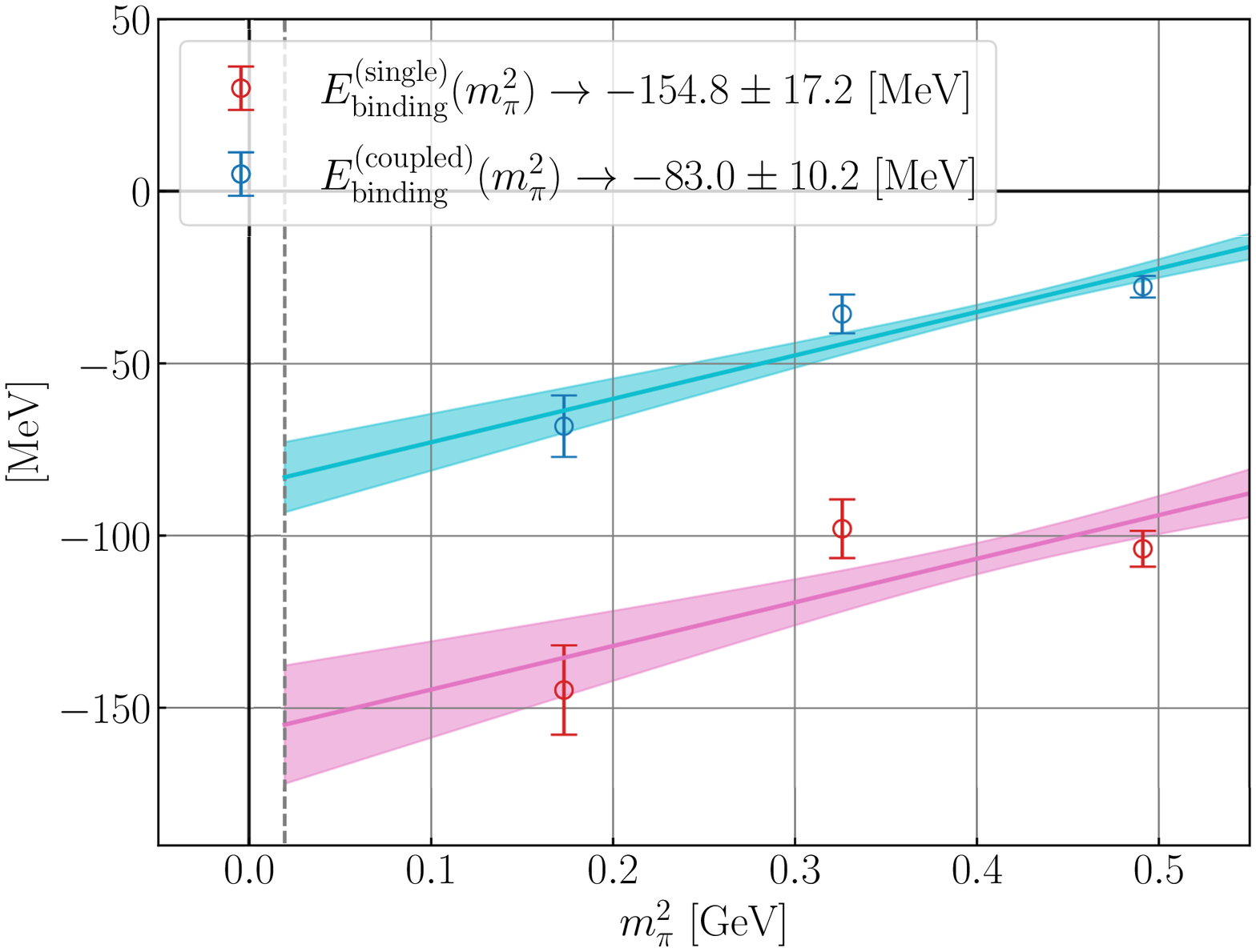}
\vskip -3cm
\caption{
The binding energy obtained by the GEM as a function of $m_\pi^2$ (open circles), together with a linear extrapolation in $m_\pi^2$ to $m_\pi=140$ MeV (solid line)  from the single channel analysis (magenta) and  the coupled-channel analysis (cyan).
\label{fig:pacscs_chiral_extrapolation_binding}}
\end{figure}

\begin{figure}[htbp]
\centering
\vskip -7cm
\includegraphics[width=0.9\linewidth]{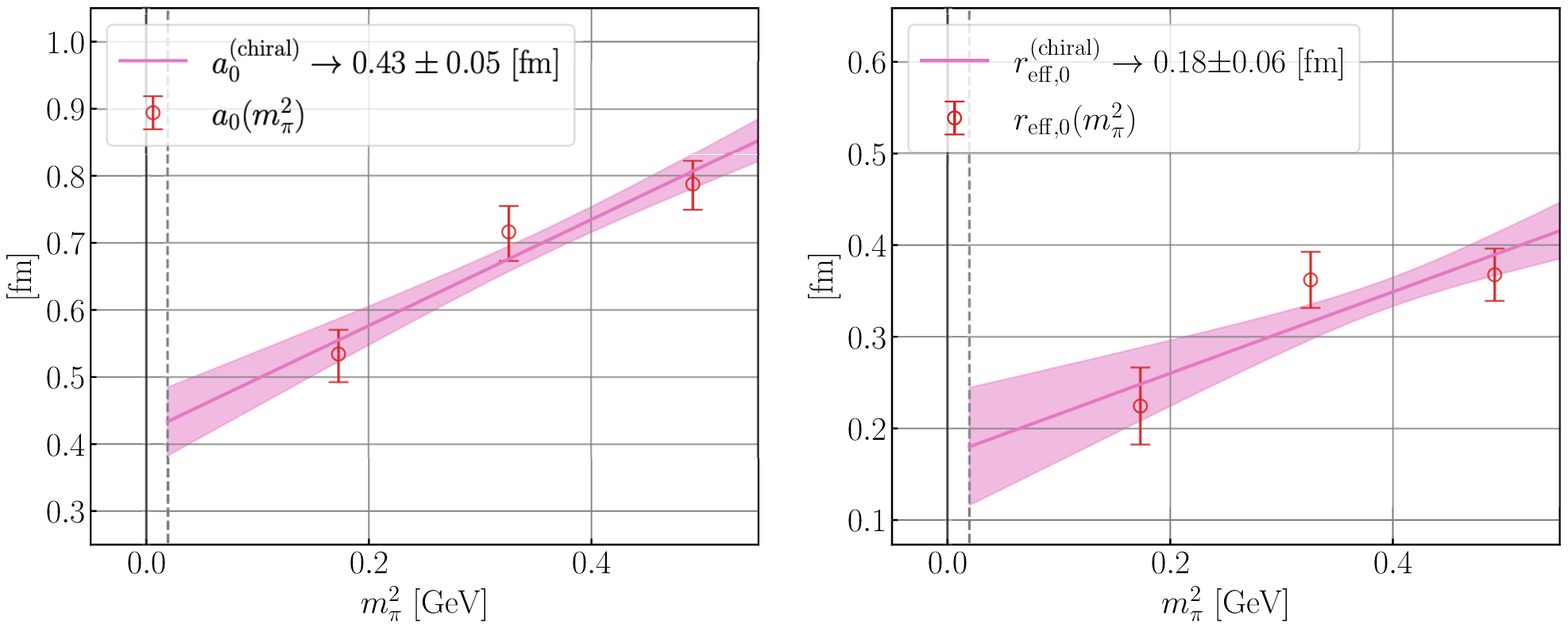}
\vskip -7cm
\caption{
ERE parameters in the coupled-channel analysis as a function of $m_\pi^2$ (open circles), 
together with a linear extrapolation in $m_\pi^2$ to $m_\pi=140$ MeV (solid line).
 (Left) The scattering length $a_0$. (Right) The effective range $r_{\rm eff,0}$. Both are defined in eq.(\ref{eq:ERE_pcotdelta}).\label{fig:pacscs_chiral_extrapolation_ERE}}
 \end{figure}

\begin{figure}[htb]
\centering
\vskip -3cm
\includegraphics[width=0.5\linewidth]{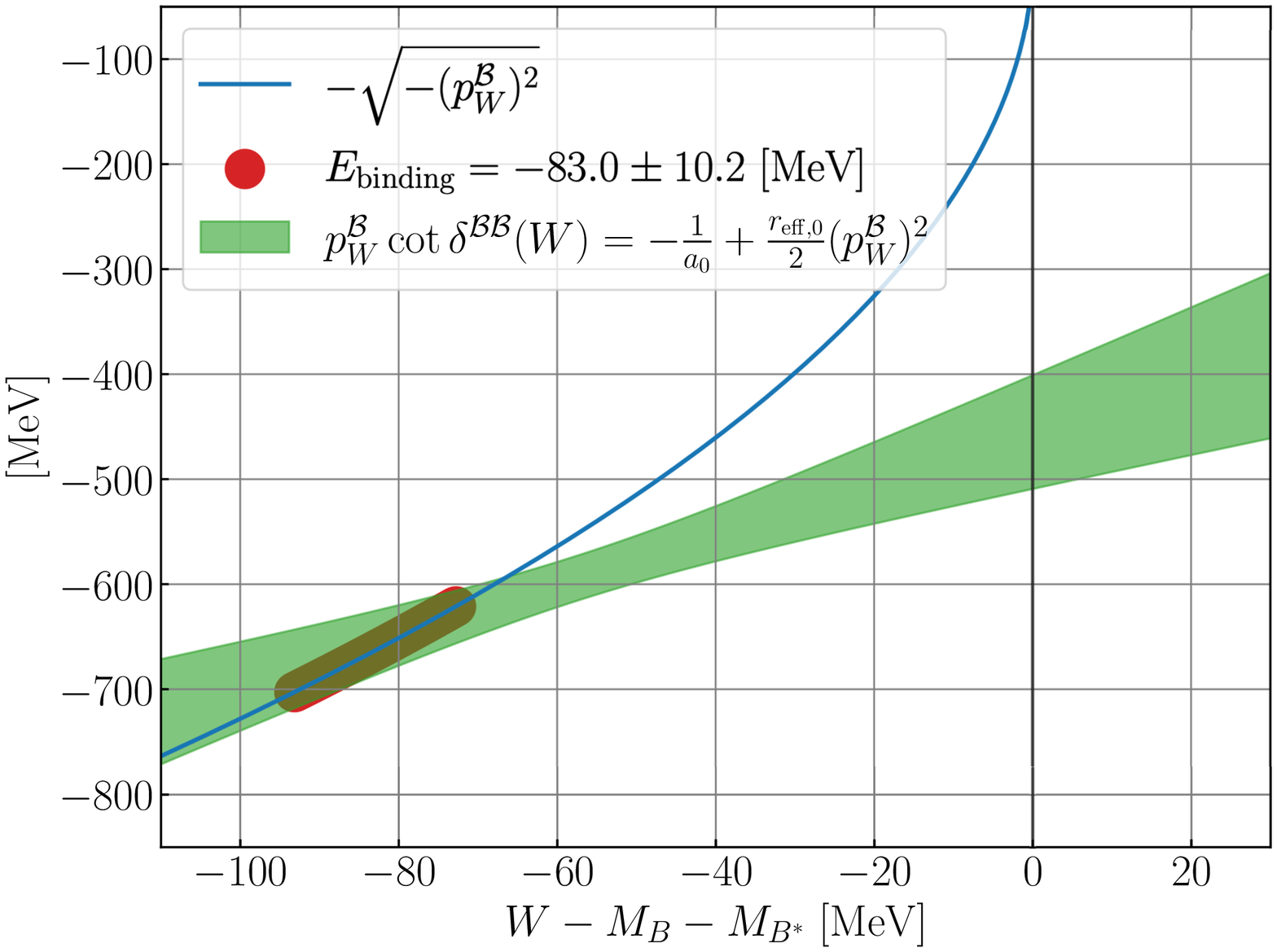}
\vskip -3cm
\caption{
The ERE at $m_\pi = 140$ MeV obtained with  $a_0$ and $r_{\rm eff,0}$ by linear extrapolations in $m_\pi^2$ (green band),
together with $-\sqrt{-p^2}$ (bleu solid line). 
An intersection of the two give a pole of the $T$-matrix, whose position is consistent with
the binding energy by the GEM at $m_\pi = 140$ MeV by a linear extrapolations in $m_\pi^2$ (red thick curve along $-\sqrt{-p^2}$).
\label{fig:pacscs_chiral_extrapolation_pcotdelta}}
\end{figure}

Fig.~\ref{fig:pacscs_LIPPvsGEM} shows scattering phase shifts as function of the energy from the $\B$ threshold ($W-m_B - m_{B^*}$)
at $m_\pi\simeq 701$ MeV (upper left), 571 MeV (upper right) and 416 MeV (lower left), 
obtained below the $\B^*$ threshold but by the coupled-channel analysis.  
Physical phase shifts $\delta$ are calculated in  the scattering region at $0< W-M_B-M_{B^*} < W - 2 M_{B^*}\simeq 45$ MeV,
while bound states are examined at $W-M_B-M_{B^*}<0$ using the analyticity of the $S$-matrix.
As mentioned before, a bound state appears at the intersection between  $p\cot\delta$ (pink,orange,red lines) and $-\sqrt{-p^2}$ (blue lines). It is observed that the system produces a pole of the $T$-matrix at each pion mass,
which satisfies a physical pole condition, eq.(\ref{eq:bound_condition}), 
so that one physical bound state exists at each pion mass. 
The thick line drawn along $-\sqrt{-p^2}$ curve is the binding energy independently obtained 
from the Schr\"odinger equation  by the Gaussian Expansion Method(GEM)\cite{Kamimura:1988zz},
which is consistent with the pole from the intersection.
Here we set a number of bases of the GEM to 50 
and the range parameters were set to be a geometric sequence with $b_1=100$ [1/fm$^2$] and $b_{50} = 0.0348$ [1/fm$^2$].

Fig.~\ref{fig:pacscs_chiral_extrapolation_binding} compares the binding energy obtained by the GEM in the coupled channel analysis (cyan)
with the one in the single channel analysis (magenta) as a function of $m_\pi^2$ (open circles),
together with a linear extrapolation in $m_\pi^2$ to the physical pion mass $m_\pi=140$ MeV (solid line),
which predicts the binding energy at the physical pion mass as
\beqa
  \label{eq:bindingenergy_phys}
  E_{\rm binding}^{\rm (single,phys)}&=& -154.8\pm 17.2 \mbox{ MeV},\quad
  E_{\rm binding}^{\rm (coupled,phys)}= -83.0\pm 10.2 \mbox{ MeV},
\eeqa
where errors are statistical only. 
A comparison of two results shows an about 40-50\%  reduction of the binding energy from the single channel analysis to the coupled channel analysis,
probably due to  large off-diagonal components of potentials.
Thus this systematics is attributed to virtual transitions such that $\B \to \B^* \to \B$,  which may easily occurs since
the $\B^*$ threshold is only 45 MeV above the $\B$ threshold.
Therefore, an inclusion of virtual $\B^*$ effect is required to predict physical observables such as the binding energy of the tetra-quark state $T_{bb}$
accurately in lattice QCD.

Fig.~\ref{fig:pacscs_chiral_extrapolation_ERE} shows the scattering length $a_0$ (Left) and the effective range $r_{\rm eff,0}$ (Right) in
the coupled channel analysis, obtained from 
the ERE fit \eqref{eq:ERE_pcotdelta},
as a function of $m_\pi^2$ (open circles), together with a linear extrapolation in $m_\pi^2$ to the physical pion mass $m_\pi=140$ MeV (solid line),
which leads to
\be
  a_0^{\rm (coupled,phys)} = 0.43\pm 0.05 \mbox{ fm},\quad
  r_{\rm eff,0}^{\rm (coupled,phys)}= 0.18\pm 0.06 \mbox{ fm},
    \label{eq:reff_phys}
\ee
at the physical pion mass, where errors are again statistical only.

In Fig.~\ref{fig:pacscs_chiral_extrapolation_pcotdelta},  the binding energy at the physical pion mass 
is alternatively estimated  from an intersection
between  $-\sqrt{-p^2}$ (blue solid line) and $p\cot\delta(p)$ (green band)  with $a_0^{\rm (coupled,phys)}$ and $r_{\rm eff,0}^{\rm (coupled,phys)}$ in \eqref{eq:reff_phys},
which not only satisfies the physical pole condition \eqref{eq:bound_condition} but also
well agrees with the binding energy by the GEM extrapolated directly to the physical pion mass (red  thick curve along  $-\sqrt{-p^2}$).
The agreement in the binding energy between the two method provide a strong support for reliability of our analysis.

\section{Conclusions\label{sec:conclusion}}
In this paper, we have extracted scattering quantities through $S$-wave potentials between $\bar B$ and $\bar B^*$ mesons with quantum numbers $I(J^P)=0(1^+)$, applying the coupled channel HAL QCD method to this single channel scattering. 
We have employed the NRQCD action for $b$ quarks to incorporate effects of their propagations in space. 
This paper presents the first analysis from a combination of the NRQCD action with the  HAL QCD method. 
Physical observables such as the binding energy, the scattering length and the effective range obtained on  (2+1)-flavor full QCD configurations at three pion masses are extrapolated to the physical pion mass.  

Since off-diagonal potentials are asymmetric and comparable in magnitude to diagonal ones, as shown in Fig.\ref{fig:potential_fit_coupled}, 
we have employed non-hermitian $2\times 2$ potentials in order to include the non-locality caused by virtual $\B^*$ states
into a single channel potential  as $U_{\rm eff}^{\B\B}$.
The single channel analysis with $U_{\rm eff}^{\B\B}$ show that the system with $\bar B$ and $\bar B^*$ mesons have a bound state corresponding to a doubly-bottom tetra-quark $T_{bb}$, whose binding energy is smaller by 40-50 \% than the one from the standard single channel analysis without non-locality.
This explicitly demonstrates an importance of virtual transitions between $\B$ and $\B^*$ channels to the tetra-quark state $T_{bb}$.
Thus it may give some hints on the nature  of the tetra-quark state $T_{bb}$ such as its internal structure.

In addition to statistical errors quoted in  eq.~\eqref{eq:bindingenergy_phys}-\eqref{eq:reff_phys},
we here estimate systematic errors in our result,
which are caused by a truncation of the NRQCD expansion, a truncation of the perturbative matching between the NRQCD hamiltonian 
and QCD, the finite lattice spacing, the finite volume and the chiral extrapolation in lattice QCD simulations, and so on.
Since these systematic errors are difficult to evaluate explicitly and precisely, we focus our attention on errors associated with the NRQCD action for $b$ quarks and employ previous studies\cite{Leskovec:2019ioa,Ishikawa:1997sx,Hashimoto:1997sr} for rough estimations.     
Effects of these systematics on the binding energy may be about 20 MeV at most, 
and other systematic errors such as  the finite lattice spacing, the finite volume and the chiral extrapolation are probably much smaller than 20 MeV,
and thus are included in this 20 MeV.
We then obtain
\beqa
  \label{eq:bindingenergy_phys_sys}
  E_{\rm binding}^{\rm (single,phys)}&=& -154.8\pm 17.2\pm 20 \mbox{ MeV},\quad
  E_{\rm binding}^{\rm (coupled,phys)}= -83.0\pm 10.2\pm 20  \mbox{ MeV}.
\eeqa
for the final estimate of the binding energy including systematic errors.
 
\begin{figure}[htb]
\centering
\includegraphics[width=0.9\linewidth]{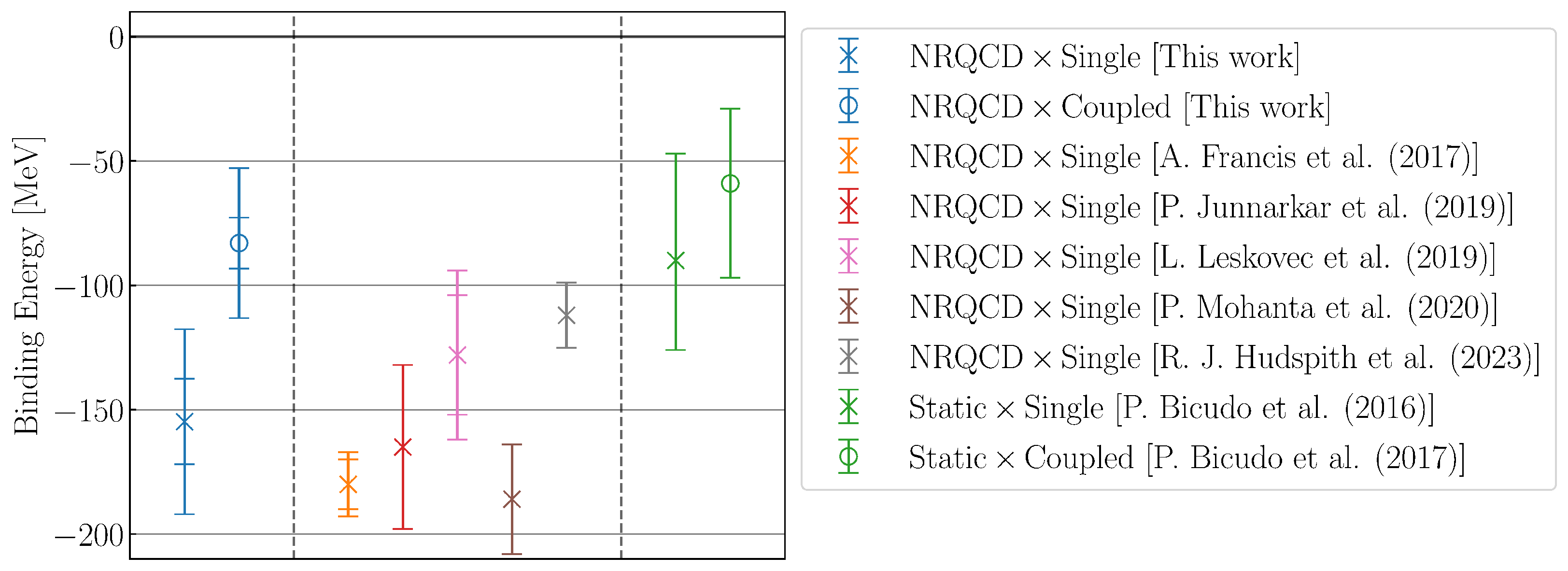}
\caption{A comparison of binding energies for the tetra-quark bound state $T_{bb}$ among several lattice QCD calculations,
from the HAL QCD potential with the NRQCD $b$ quark (blue), 
spectra with the NRQCD $b$ quark (orange, red, magenta and brown), and
the static quark potential with the static $b$ quark (green).
Crosses indicate results from the single-channel analysis, while open circles from the coupled-channel analysis.\label{fig:comparison_with_prev_studies}}
\end{figure}

 We compare  these final results with latest lattice studies\cite{Leskovec:2019ioa,Mohanta:2020eed,Francis:2016hui,Junnarkar:2018twb,Bicudo:2015kna,Bicudo:2016ooe,Hudspith:2023loy} in Fig.~\ref{fig:comparison_with_prev_studies}.  
From the comparison, we draw following conclusions.
\begin{itemize}
\item Within the same combination, the NRQCD $b$ quark and the single-channel analysis,
our result (blue cross) roughly agrees with others (orange cross and red cross) within large errors.
\item Within the  single-channel analysis, the binding energy is a little larger for our result with the NRQCD $b$ quark (blue cross)
than the one with the static $b$ quark (green cross).
\item Our result with the NRQCD $b$ quark in  the coupled-channel analysis (blue open circle) is new.  
The result shows that
a reduction of the binding energy from the single-channel analysis (blue cross) to the  the coupled-channel analysis (blue open circle), about
70 MeV,  is much larger than the reduction with the static $b$ quark ( green cross and open circle).
\end{itemize}
While an existence a tetra-quark bound state $T_{bb}$ is a robust prediction in lattice QCD, 
reductions of systematic errors will be needed to evaluate its binding energy more precisely in future studies.

\subsection*{Acknowledgements}
We thank the PACS-CS Collaboration~\cite{PACS-CS:2008bkb} and ILDG/JLDG~\cite{Amagasa:2015zwb} for providing us their gauge configurations. 
We have used lattice QCD code of Bridge++\cite{Ueda:2014rya},
gauge-fixing code of cuLGT\cite{Vogt:2014hea} and HAL QCD analysis is based on the code written by Dr. T. Miyamoto. 
In particular, we would like to thank Dr. H. Matsufuru, who has ported NRQCD code ported into the Bridge++.
Our numerical calculation has been performed on Yukawa-21 at Yukawa Institute for Theoretical Physics (YITP) in Kyoto University. 
We thank other members of the HAL Collaboration, especially Dr. Y. Akahoshi,  Mr. K. Murakami, and Dr. H. Nemura  for fruitful discussions.
This work  is supported in part by the JSPS Grant-in-Aid  for Scientific Research (Nos. JP16H03978, JP18H05236, JP18K03628).

\bibliography{bbud01}

\begin{thebibliography}{38}%
\makeatletter
\providecommand \@ifxundefined [1]{%
 \@ifx{#1\undefined}
}%
\providecommand \@ifnum [1]{%
 \ifnum #1\expandafter \@firstoftwo
 \else \expandafter \@secondoftwo
 \fi
}%
\providecommand \@ifx [1]{%
 \ifx #1\expandafter \@firstoftwo
 \else \expandafter \@secondoftwo
 \fi
}%
\providecommand \natexlab [1]{#1}%
\providecommand \enquote  [1]{``#1''}%
\providecommand \bibnamefont  [1]{#1}%
\providecommand \bibfnamefont [1]{#1}%
\providecommand \citenamefont [1]{#1}%
\providecommand \href@noop [0]{\@secondoftwo}%
\providecommand \href [0]{\begingroup \@sanitize@url \@href}%
\providecommand \@href[1]{\@@startlink{#1}\@@href}%
\providecommand \@@href[1]{\endgroup#1\@@endlink}%
\providecommand \@sanitize@url [0]{\catcode `\\12\catcode `\$12\catcode
  `\&12\catcode `\#12\catcode `\^12\catcode `\_12\catcode `\%12\relax}%
\providecommand \@@startlink[1]{}%
\providecommand \@@endlink[0]{}%
\providecommand \url  [0]{\begingroup\@sanitize@url \@url }%
\providecommand \@url [1]{\endgroup\@href {#1}{\urlprefix }}%
\providecommand \urlprefix  [0]{URL }%
\providecommand \Eprint [0]{\href }%
\providecommand \doibase [0]{http://dx.doi.org/}%
\providecommand \selectlanguage [0]{\@gobble}%
\providecommand \bibinfo  [0]{\@secondoftwo}%
\providecommand \bibfield  [0]{\@secondoftwo}%
\providecommand \translation [1]{[#1]}%
\providecommand \BibitemOpen [0]{}%
\providecommand \bibitemStop [0]{}%
\providecommand \bibitemNoStop [0]{.\EOS\space}%
\providecommand \EOS [0]{\spacefactor3000\relax}%
\providecommand \BibitemShut  [1]{\csname bibitem#1\endcsname}%
\let\auto@bib@innerbib\@empty
\bibitem [{\citenamefont {Aaij}\ \emph {et~al.}(2020)\citenamefont {Aaij} \emph
  {et~al.}}]{LHCb:2020pxc}%
  \BibitemOpen
  \bibfield  {author} {\bibinfo {author} {\bibfnamefont {R.}~\bibnamefont
  {Aaij}} \emph {et~al.} (\bibinfo {collaboration} {LHCb}),\ }\href {\doibase
  10.1103/PhysRevD.102.112003} {\bibfield  {journal} {\bibinfo  {journal}
  {Phys. Rev. D}\ }\textbf {\bibinfo {volume} {102}},\ \bibinfo {pages}
  {112003} (\bibinfo {year} {2020})},\ \Eprint
  {http://arxiv.org/abs/2009.00026} {arXiv:2009.00026 [hep-ex]} \BibitemShut
  {NoStop}%
\bibitem [{\citenamefont {Aaij}\ \emph {et~al.}(2021)\citenamefont {Aaij} \emph
  {et~al.}}]{LHCb:2021vvq}%
  \BibitemOpen
  \bibfield  {author} {\bibinfo {author} {\bibfnamefont {R.}~\bibnamefont
  {Aaij}} \emph {et~al.} (\bibinfo {collaboration} {LHCb}),\ }\href {\doibase
  10.1038/s41567-022-01614-y} {\  (\bibinfo {year} {2021}),\
  10.1038/s41567-022-01614-y},\ \Eprint {http://arxiv.org/abs/2109.01038}
  {arXiv:2109.01038 [hep-ex]} \BibitemShut {NoStop}%
\bibitem [{\citenamefont {Aaij}\ \emph {et~al.}(2015)\citenamefont {Aaij} \emph
  {et~al.}}]{LHCb:2015yax}%
  \BibitemOpen
  \bibfield  {author} {\bibinfo {author} {\bibfnamefont {R.}~\bibnamefont
  {Aaij}} \emph {et~al.} (\bibinfo {collaboration} {LHCb}),\ }\href {\doibase
  10.1103/PhysRevLett.115.072001} {\bibfield  {journal} {\bibinfo  {journal}
  {Phys. Rev. Lett.}\ }\textbf {\bibinfo {volume} {115}},\ \bibinfo {pages}
  {072001} (\bibinfo {year} {2015})},\ \Eprint
  {http://arxiv.org/abs/1507.03414} {arXiv:1507.03414 [hep-ex]} \BibitemShut
  {NoStop}%
\bibitem [{\citenamefont {Bondar}\ \emph {et~al.}(2012)\citenamefont {Bondar}
  \emph {et~al.}}]{Belle:2011aa}%
  \BibitemOpen
  \bibfield  {author} {\bibinfo {author} {\bibfnamefont {A.}~\bibnamefont
  {Bondar}} \emph {et~al.} (\bibinfo {collaboration} {Belle}),\ }\href
  {\doibase 10.1103/PhysRevLett.108.122001} {\bibfield  {journal} {\bibinfo
  {journal} {Phys. Rev. Lett.}\ }\textbf {\bibinfo {volume} {108}},\ \bibinfo
  {pages} {122001} (\bibinfo {year} {2012})},\ \Eprint
  {http://arxiv.org/abs/1110.2251} {arXiv:1110.2251 [hep-ex]} \BibitemShut
  {NoStop}%
\bibitem [{\citenamefont {Cheng}\ \emph {et~al.}(2021)\citenamefont {Cheng},
  \citenamefont {Li}, \citenamefont {Liu}, \citenamefont {Si},\ and\
  \citenamefont {Yao}}]{Cheng:2020wxa}%
  \BibitemOpen
  \bibfield  {author} {\bibinfo {author} {\bibfnamefont {J.-B.}\ \bibnamefont
  {Cheng}}, \bibinfo {author} {\bibfnamefont {S.-Y.}\ \bibnamefont {Li}},
  \bibinfo {author} {\bibfnamefont {Y.-R.}\ \bibnamefont {Liu}}, \bibinfo
  {author} {\bibfnamefont {Z.-G.}\ \bibnamefont {Si}}, \ and\ \bibinfo {author}
  {\bibfnamefont {T.}~\bibnamefont {Yao}},\ }\href {\doibase
  10.1088/1674-1137/abde2f} {\bibfield  {journal} {\bibinfo  {journal} {Chin.
  Phys. C}\ }\textbf {\bibinfo {volume} {45}},\ \bibinfo {pages} {043102}
  (\bibinfo {year} {2021})},\ \Eprint {http://arxiv.org/abs/2008.00737}
  {arXiv:2008.00737 [hep-ph]} \BibitemShut {NoStop}%
\bibitem [{\citenamefont {Karliner}\ and\ \citenamefont
  {Rosner}(2017)}]{Karliner:2017qjm}%
  \BibitemOpen
  \bibfield  {author} {\bibinfo {author} {\bibfnamefont {M.}~\bibnamefont
  {Karliner}}\ and\ \bibinfo {author} {\bibfnamefont {J.~L.}\ \bibnamefont
  {Rosner}},\ }\href {\doibase 10.1103/PhysRevLett.119.202001} {\bibfield
  {journal} {\bibinfo  {journal} {Phys. Rev. Lett.}\ }\textbf {\bibinfo
  {volume} {119}},\ \bibinfo {pages} {202001} (\bibinfo {year} {2017})},\
  \Eprint {http://arxiv.org/abs/1707.07666} {arXiv:1707.07666 [hep-ph]}
  \BibitemShut {NoStop}%
\bibitem [{\citenamefont {Eichten}\ and\ \citenamefont
  {Quigg}(2017)}]{Eichten:2017ffp}%
  \BibitemOpen
  \bibfield  {author} {\bibinfo {author} {\bibfnamefont {E.~J.}\ \bibnamefont
  {Eichten}}\ and\ \bibinfo {author} {\bibfnamefont {C.}~\bibnamefont
  {Quigg}},\ }\href {\doibase 10.1103/PhysRevLett.119.202002} {\bibfield
  {journal} {\bibinfo  {journal} {Phys. Rev. Lett.}\ }\textbf {\bibinfo
  {volume} {119}},\ \bibinfo {pages} {202002} (\bibinfo {year} {2017})},\
  \Eprint {http://arxiv.org/abs/1707.09575} {arXiv:1707.09575 [hep-ph]}
  \BibitemShut {NoStop}%
\bibitem [{\citenamefont {Bicudo}\ \emph {et~al.}(2016)\citenamefont {Bicudo},
  \citenamefont {Cichy}, \citenamefont {Peters},\ and\ \citenamefont
  {Wagner}}]{Bicudo:2015kna}%
  \BibitemOpen
  \bibfield  {author} {\bibinfo {author} {\bibfnamefont {P.}~\bibnamefont
  {Bicudo}}, \bibinfo {author} {\bibfnamefont {K.}~\bibnamefont {Cichy}},
  \bibinfo {author} {\bibfnamefont {A.}~\bibnamefont {Peters}}, \ and\ \bibinfo
  {author} {\bibfnamefont {M.}~\bibnamefont {Wagner}},\ }\href {\doibase
  10.1103/PhysRevD.93.034501} {\bibfield  {journal} {\bibinfo  {journal} {Phys.
  Rev. D}\ }\textbf {\bibinfo {volume} {93}},\ \bibinfo {pages} {034501}
  (\bibinfo {year} {2016})},\ \Eprint {http://arxiv.org/abs/1510.03441}
  {arXiv:1510.03441 [hep-lat]} \BibitemShut {NoStop}%
\bibitem [{\citenamefont {Bicudo}\ \emph {et~al.}(2017)\citenamefont {Bicudo},
  \citenamefont {Scheunert},\ and\ \citenamefont {Wagner}}]{Bicudo:2016ooe}%
  \BibitemOpen
  \bibfield  {author} {\bibinfo {author} {\bibfnamefont {P.}~\bibnamefont
  {Bicudo}}, \bibinfo {author} {\bibfnamefont {J.}~\bibnamefont {Scheunert}}, \
  and\ \bibinfo {author} {\bibfnamefont {M.}~\bibnamefont {Wagner}},\ }\href
  {\doibase 10.1103/PhysRevD.95.034502} {\bibfield  {journal} {\bibinfo
  {journal} {Phys. Rev. D}\ }\textbf {\bibinfo {volume} {95}},\ \bibinfo
  {pages} {034502} (\bibinfo {year} {2017})},\ \Eprint
  {http://arxiv.org/abs/1612.02758} {arXiv:1612.02758 [hep-lat]} \BibitemShut
  {NoStop}%
\bibitem [{\citenamefont {Francis}\ \emph {et~al.}(2017)\citenamefont
  {Francis}, \citenamefont {Hudspith}, \citenamefont {Lewis},\ and\
  \citenamefont {Maltman}}]{Francis:2016hui}%
  \BibitemOpen
  \bibfield  {author} {\bibinfo {author} {\bibfnamefont {A.}~\bibnamefont
  {Francis}}, \bibinfo {author} {\bibfnamefont {R.~J.}\ \bibnamefont
  {Hudspith}}, \bibinfo {author} {\bibfnamefont {R.}~\bibnamefont {Lewis}}, \
  and\ \bibinfo {author} {\bibfnamefont {K.}~\bibnamefont {Maltman}},\ }\href
  {\doibase 10.1103/PhysRevLett.118.142001} {\bibfield  {journal} {\bibinfo
  {journal} {Phys. Rev. Lett.}\ }\textbf {\bibinfo {volume} {118}},\ \bibinfo
  {pages} {142001} (\bibinfo {year} {2017})},\ \Eprint
  {http://arxiv.org/abs/1607.05214} {arXiv:1607.05214 [hep-lat]} \BibitemShut
  {NoStop}%
\bibitem [{\citenamefont {Junnarkar}\ \emph {et~al.}(2019)\citenamefont
  {Junnarkar}, \citenamefont {Mathur},\ and\ \citenamefont
  {Padmanath}}]{Junnarkar:2018twb}%
  \BibitemOpen
  \bibfield  {author} {\bibinfo {author} {\bibfnamefont {P.}~\bibnamefont
  {Junnarkar}}, \bibinfo {author} {\bibfnamefont {N.}~\bibnamefont {Mathur}}, \
  and\ \bibinfo {author} {\bibfnamefont {M.}~\bibnamefont {Padmanath}},\ }\href
  {\doibase 10.1103/PhysRevD.99.034507} {\bibfield  {journal} {\bibinfo
  {journal} {Phys. Rev. D}\ }\textbf {\bibinfo {volume} {99}},\ \bibinfo
  {pages} {034507} (\bibinfo {year} {2019})},\ \Eprint
  {http://arxiv.org/abs/1810.12285} {arXiv:1810.12285 [hep-lat]} \BibitemShut
  {NoStop}%
\bibitem [{\citenamefont {Leskovec}\ \emph {et~al.}(2019)\citenamefont
  {Leskovec}, \citenamefont {Meinel}, \citenamefont {Pflaumer},\ and\
  \citenamefont {Wagner}}]{Leskovec:2019ioa}%
  \BibitemOpen
  \bibfield  {author} {\bibinfo {author} {\bibfnamefont {L.}~\bibnamefont
  {Leskovec}}, \bibinfo {author} {\bibfnamefont {S.}~\bibnamefont {Meinel}},
  \bibinfo {author} {\bibfnamefont {M.}~\bibnamefont {Pflaumer}}, \ and\
  \bibinfo {author} {\bibfnamefont {M.}~\bibnamefont {Wagner}},\ }\href
  {\doibase 10.1103/PhysRevD.100.014503} {\bibfield  {journal} {\bibinfo
  {journal} {Phys. Rev. D}\ }\textbf {\bibinfo {volume} {100}},\ \bibinfo
  {pages} {014503} (\bibinfo {year} {2019})},\ \Eprint
  {http://arxiv.org/abs/1904.04197} {arXiv:1904.04197 [hep-lat]} \BibitemShut
  {NoStop}%
\bibitem [{\citenamefont {Mohanta}\ and\ \citenamefont
  {Basak}(2020)}]{Mohanta:2020eed}%
  \BibitemOpen
  \bibfield  {author} {\bibinfo {author} {\bibfnamefont {P.}~\bibnamefont
  {Mohanta}}\ and\ \bibinfo {author} {\bibfnamefont {S.}~\bibnamefont
  {Basak}},\ }\href {\doibase 10.1103/PhysRevD.102.094516} {\bibfield
  {journal} {\bibinfo  {journal} {Phys. Rev. D}\ }\textbf {\bibinfo {volume}
  {102}},\ \bibinfo {pages} {094516} (\bibinfo {year} {2020})},\ \Eprint
  {http://arxiv.org/abs/2008.11146} {arXiv:2008.11146 [hep-lat]} \BibitemShut
  {NoStop}%
\bibitem [{\citenamefont {Hudspith}\ and\ \citenamefont
  {Mohler}(2023)}]{Hudspith:2023loy}%
  \BibitemOpen
  \bibfield  {author} {\bibinfo {author} {\bibfnamefont {R.~J.}\ \bibnamefont
  {Hudspith}}\ and\ \bibinfo {author} {\bibfnamefont {D.}~\bibnamefont
  {Mohler}},\ }\href@noop {} {\  (\bibinfo {year} {2023})},\ \Eprint
  {http://arxiv.org/abs/2303.17295} {arXiv:2303.17295 [hep-lat]} \BibitemShut
  {NoStop}%
\bibitem [{\citenamefont {Luscher}(1991)}]{Luscher:1990ux}%
  \BibitemOpen
  \bibfield  {author} {\bibinfo {author} {\bibfnamefont {M.}~\bibnamefont
  {Luscher}},\ }\href {\doibase 10.1016/0550-3213(91)90366-6} {\bibfield
  {journal} {\bibinfo  {journal} {Nucl. Phys. B}\ }\textbf {\bibinfo {volume}
  {354}},\ \bibinfo {pages} {531} (\bibinfo {year} {1991})}\BibitemShut
  {NoStop}%
\bibitem [{\citenamefont {Aoki}\ \emph
  {et~al.}(2013{\natexlab{a}})\citenamefont {Aoki}, \citenamefont {Ishii},
  \citenamefont {Doi}, \citenamefont {Ikeda},\ and\ \citenamefont
  {Inoue}}]{Aoki:2013cra}%
  \BibitemOpen
  \bibfield  {author} {\bibinfo {author} {\bibfnamefont {S.}~\bibnamefont
  {Aoki}}, \bibinfo {author} {\bibfnamefont {N.}~\bibnamefont {Ishii}},
  \bibinfo {author} {\bibfnamefont {T.}~\bibnamefont {Doi}}, \bibinfo {author}
  {\bibfnamefont {Y.}~\bibnamefont {Ikeda}}, \ and\ \bibinfo {author}
  {\bibfnamefont {T.}~\bibnamefont {Inoue}},\ }\href {\doibase
  10.1103/PhysRevD.88.014036} {\bibfield  {journal} {\bibinfo  {journal} {Phys.
  Rev. D}\ }\textbf {\bibinfo {volume} {88}},\ \bibinfo {pages} {014036}
  (\bibinfo {year} {2013}{\natexlab{a}})},\ \Eprint
  {http://arxiv.org/abs/1303.2210} {arXiv:1303.2210 [hep-lat]} \BibitemShut
  {NoStop}%
\bibitem [{\citenamefont {Ishii}\ \emph {et~al.}(2007)\citenamefont {Ishii},
  \citenamefont {Aoki},\ and\ \citenamefont {Hatsuda}}]{Ishii:2006ec}%
  \BibitemOpen
  \bibfield  {author} {\bibinfo {author} {\bibfnamefont {N.}~\bibnamefont
  {Ishii}}, \bibinfo {author} {\bibfnamefont {S.}~\bibnamefont {Aoki}}, \ and\
  \bibinfo {author} {\bibfnamefont {T.}~\bibnamefont {Hatsuda}},\ }\href
  {\doibase 10.1103/PhysRevLett.99.022001} {\bibfield  {journal} {\bibinfo
  {journal} {Phys. Rev. Lett.}\ }\textbf {\bibinfo {volume} {99}},\ \bibinfo
  {pages} {022001} (\bibinfo {year} {2007})},\ \Eprint
  {http://arxiv.org/abs/nucl-th/0611096} {arXiv:nucl-th/0611096} \BibitemShut
  {NoStop}%
\bibitem [{\citenamefont {Aoki}\ \emph {et~al.}(2010)\citenamefont {Aoki},
  \citenamefont {Hatsuda},\ and\ \citenamefont {Ishii}}]{Aoki:2009ji}%
  \BibitemOpen
  \bibfield  {author} {\bibinfo {author} {\bibfnamefont {S.}~\bibnamefont
  {Aoki}}, \bibinfo {author} {\bibfnamefont {T.}~\bibnamefont {Hatsuda}}, \
  and\ \bibinfo {author} {\bibfnamefont {N.}~\bibnamefont {Ishii}},\ }\href
  {\doibase 10.1143/PTP.123.89} {\bibfield  {journal} {\bibinfo  {journal}
  {Prog. Theor. Phys.}\ }\textbf {\bibinfo {volume} {123}},\ \bibinfo {pages}
  {89} (\bibinfo {year} {2010})},\ \Eprint {http://arxiv.org/abs/0909.5585}
  {arXiv:0909.5585 [hep-lat]} \BibitemShut {NoStop}%
\bibitem [{\citenamefont {Aoki}\ \emph {et~al.}(2012)\citenamefont {Aoki},
  \citenamefont {Doi}, \citenamefont {Hatsuda}, \citenamefont {Ikeda},
  \citenamefont {Inoue}, \citenamefont {Ishii}, \citenamefont {Murano},
  \citenamefont {Nemura},\ and\ \citenamefont {Sasaki}}]{Aoki:2012tk}%
  \BibitemOpen
  \bibfield  {author} {\bibinfo {author} {\bibfnamefont {S.}~\bibnamefont
  {Aoki}}, \bibinfo {author} {\bibfnamefont {T.}~\bibnamefont {Doi}}, \bibinfo
  {author} {\bibfnamefont {T.}~\bibnamefont {Hatsuda}}, \bibinfo {author}
  {\bibfnamefont {Y.}~\bibnamefont {Ikeda}}, \bibinfo {author} {\bibfnamefont
  {T.}~\bibnamefont {Inoue}}, \bibinfo {author} {\bibfnamefont
  {N.}~\bibnamefont {Ishii}}, \bibinfo {author} {\bibfnamefont
  {K.}~\bibnamefont {Murano}}, \bibinfo {author} {\bibfnamefont
  {H.}~\bibnamefont {Nemura}}, \ and\ \bibinfo {author} {\bibfnamefont
  {K.}~\bibnamefont {Sasaki}} (\bibinfo {collaboration} {HAL QCD}),\ }\href
  {\doibase 10.1093/ptep/pts010} {\bibfield  {journal} {\bibinfo  {journal}
  {PTEP}\ }\textbf {\bibinfo {volume} {2012}},\ \bibinfo {pages} {01A105}
  (\bibinfo {year} {2012})},\ \Eprint {http://arxiv.org/abs/1206.5088}
  {arXiv:1206.5088 [hep-lat]} \BibitemShut {NoStop}%
\bibitem [{\citenamefont {Aoki}\ \emph {et~al.}(2011)\citenamefont {Aoki},
  \citenamefont {Ishii}, \citenamefont {Doi}, \citenamefont {Hatsuda},
  \citenamefont {Ikeda}, \citenamefont {Inoue}, \citenamefont {Murano},
  \citenamefont {Nemura},\ and\ \citenamefont {Sasaki}}]{Aoki:2011gt}%
  \BibitemOpen
  \bibfield  {author} {\bibinfo {author} {\bibfnamefont {S.}~\bibnamefont
  {Aoki}}, \bibinfo {author} {\bibfnamefont {N.}~\bibnamefont {Ishii}},
  \bibinfo {author} {\bibfnamefont {T.}~\bibnamefont {Doi}}, \bibinfo {author}
  {\bibfnamefont {T.}~\bibnamefont {Hatsuda}}, \bibinfo {author} {\bibfnamefont
  {Y.}~\bibnamefont {Ikeda}}, \bibinfo {author} {\bibfnamefont
  {T.}~\bibnamefont {Inoue}}, \bibinfo {author} {\bibfnamefont
  {K.}~\bibnamefont {Murano}}, \bibinfo {author} {\bibfnamefont
  {H.}~\bibnamefont {Nemura}}, \ and\ \bibinfo {author} {\bibfnamefont
  {K.}~\bibnamefont {Sasaki}} (\bibinfo {collaboration} {HAL QCD}),\ }\href
  {\doibase 10.2183/pjab.87.509} {\bibfield  {journal} {\bibinfo  {journal}
  {Proc. Japan Acad. B}\ }\textbf {\bibinfo {volume} {87}},\ \bibinfo {pages}
  {509} (\bibinfo {year} {2011})},\ \Eprint {http://arxiv.org/abs/1106.2281}
  {arXiv:1106.2281 [hep-lat]} \BibitemShut {NoStop}%
\bibitem [{\citenamefont {Aoki}\ \emph
  {et~al.}(2013{\natexlab{b}})\citenamefont {Aoki}, \citenamefont {Charron},
  \citenamefont {Doi}, \citenamefont {Hatsuda}, \citenamefont {Inoue},\ and\
  \citenamefont {Ishii}}]{Aoki:2012bb}%
  \BibitemOpen
  \bibfield  {author} {\bibinfo {author} {\bibfnamefont {S.}~\bibnamefont
  {Aoki}}, \bibinfo {author} {\bibfnamefont {B.}~\bibnamefont {Charron}},
  \bibinfo {author} {\bibfnamefont {T.}~\bibnamefont {Doi}}, \bibinfo {author}
  {\bibfnamefont {T.}~\bibnamefont {Hatsuda}}, \bibinfo {author} {\bibfnamefont
  {T.}~\bibnamefont {Inoue}}, \ and\ \bibinfo {author} {\bibfnamefont
  {N.}~\bibnamefont {Ishii}},\ }\href {\doibase 10.1103/PhysRevD.87.034512}
  {\bibfield  {journal} {\bibinfo  {journal} {Phys. Rev. D}\ }\textbf {\bibinfo
  {volume} {87}},\ \bibinfo {pages} {034512} (\bibinfo {year}
  {2013}{\natexlab{b}})},\ \Eprint {http://arxiv.org/abs/1212.4896}
  {arXiv:1212.4896 [hep-lat]} \BibitemShut {NoStop}%
\bibitem [{\citenamefont {Ishii}\ \emph {et~al.}(2012)\citenamefont {Ishii},
  \citenamefont {Aoki}, \citenamefont {Doi}, \citenamefont {Hatsuda},
  \citenamefont {Ikeda}, \citenamefont {Inoue}, \citenamefont {Murano},
  \citenamefont {Nemura},\ and\ \citenamefont {Sasaki}}]{Ishii:2012ssm}%
  \BibitemOpen
  \bibfield  {author} {\bibinfo {author} {\bibfnamefont {N.}~\bibnamefont
  {Ishii}}, \bibinfo {author} {\bibfnamefont {S.}~\bibnamefont {Aoki}},
  \bibinfo {author} {\bibfnamefont {T.}~\bibnamefont {Doi}}, \bibinfo {author}
  {\bibfnamefont {T.}~\bibnamefont {Hatsuda}}, \bibinfo {author} {\bibfnamefont
  {Y.}~\bibnamefont {Ikeda}}, \bibinfo {author} {\bibfnamefont
  {T.}~\bibnamefont {Inoue}}, \bibinfo {author} {\bibfnamefont
  {K.}~\bibnamefont {Murano}}, \bibinfo {author} {\bibfnamefont
  {H.}~\bibnamefont {Nemura}}, \ and\ \bibinfo {author} {\bibfnamefont
  {K.}~\bibnamefont {Sasaki}} (\bibinfo {collaboration} {HAL QCD}),\ }\href
  {\doibase 10.1016/j.physletb.2012.04.076} {\bibfield  {journal} {\bibinfo
  {journal} {Phys. Lett. B}\ }\textbf {\bibinfo {volume} {712}},\ \bibinfo
  {pages} {437} (\bibinfo {year} {2012})},\ \Eprint
  {http://arxiv.org/abs/1203.3642} {arXiv:1203.3642 [hep-lat]} \BibitemShut
  {NoStop}%
\bibitem [{\citenamefont {Taylor}()}]{taylorQuantumTheoryNonrelativistic1972}%
  \BibitemOpen
  \bibfield  {author} {\bibinfo {author} {\bibfnamefont {J.~R.}\ \bibnamefont
  {Taylor}},\ }\href@noop {} {\emph {\bibinfo {title} {The {{Quantum Theory}}
  of {{Nonrelativistic Collisions}}}}}\ (\bibinfo  {publisher} {{Dover
  books}})\BibitemShut {NoStop}%
\bibitem [{\citenamefont {Zyla}\ \emph {et~al.}(2020)\citenamefont {Zyla} \emph
  {et~al.}}]{ParticleDataGroup:2020ssz}%
  \BibitemOpen
  \bibfield  {author} {\bibinfo {author} {\bibfnamefont {P.~A.}\ \bibnamefont
  {Zyla}} \emph {et~al.} (\bibinfo {collaboration} {Particle Data Group}),\
  }\href {\doibase 10.1093/ptep/ptaa104} {\bibfield  {journal} {\bibinfo
  {journal} {PTEP}\ }\textbf {\bibinfo {volume} {2020}},\ \bibinfo {pages}
  {083C01} (\bibinfo {year} {2020})}\BibitemShut {NoStop}%
\bibitem [{\citenamefont {Thacker}\ and\ \citenamefont
  {Lepage}(1991)}]{Thacker:1990bm}%
  \BibitemOpen
  \bibfield  {author} {\bibinfo {author} {\bibfnamefont {B.~A.}\ \bibnamefont
  {Thacker}}\ and\ \bibinfo {author} {\bibfnamefont {G.~P.}\ \bibnamefont
  {Lepage}},\ }\href {\doibase 10.1103/PhysRevD.43.196} {\bibfield  {journal}
  {\bibinfo  {journal} {Phys. Rev. D}\ }\textbf {\bibinfo {volume} {43}},\
  \bibinfo {pages} {196} (\bibinfo {year} {1991})}\BibitemShut {NoStop}%
\bibitem [{\citenamefont {Foldy}\ and\ \citenamefont
  {Wouthuysen}(1950)}]{Foldy:1949wa}%
  \BibitemOpen
  \bibfield  {author} {\bibinfo {author} {\bibfnamefont {L.~L.}\ \bibnamefont
  {Foldy}}\ and\ \bibinfo {author} {\bibfnamefont {S.~A.}\ \bibnamefont
  {Wouthuysen}},\ }\href {\doibase 10.1103/PhysRev.78.29} {\bibfield  {journal}
  {\bibinfo  {journal} {Phys. Rev.}\ }\textbf {\bibinfo {volume} {78}},\
  \bibinfo {pages} {29} (\bibinfo {year} {1950})}\BibitemShut {NoStop}%
\bibitem [{\citenamefont {Tani}()}]{taniConnectionParticleModels1951}%
  \BibitemOpen
  \bibfield  {author} {\bibinfo {author} {\bibfnamefont {S.}~\bibnamefont
  {Tani}},\ }\href {\doibase 10.1143/ptp/6.3.267} {\bibfield  {journal}
  {\bibinfo  {journal} {Prog. Theor. Phys.}\ }\textbf {\bibinfo {volume} {6}},\
  \bibinfo {pages} {267}}\BibitemShut {NoStop}%
\bibitem [{\citenamefont {Lepage}\ \emph {et~al.}(1992)\citenamefont {Lepage},
  \citenamefont {Magnea}, \citenamefont {Nakhleh}, \citenamefont {Magnea},\
  and\ \citenamefont {Hornbostel}}]{Lepage:1992tx}%
  \BibitemOpen
  \bibfield  {author} {\bibinfo {author} {\bibfnamefont {G.~P.}\ \bibnamefont
  {Lepage}}, \bibinfo {author} {\bibfnamefont {L.}~\bibnamefont {Magnea}},
  \bibinfo {author} {\bibfnamefont {C.}~\bibnamefont {Nakhleh}}, \bibinfo
  {author} {\bibfnamefont {U.}~\bibnamefont {Magnea}}, \ and\ \bibinfo {author}
  {\bibfnamefont {K.}~\bibnamefont {Hornbostel}},\ }\href {\doibase
  10.1103/PhysRevD.46.4052} {\bibfield  {journal} {\bibinfo  {journal} {Phys.
  Rev. D}\ }\textbf {\bibinfo {volume} {46}},\ \bibinfo {pages} {4052}
  (\bibinfo {year} {1992})},\ \Eprint {http://arxiv.org/abs/hep-lat/9205007}
  {arXiv:hep-lat/9205007} \BibitemShut {NoStop}%
\bibitem [{\citenamefont {Ishikawa}\ \emph {et~al.}(1997)\citenamefont
  {Ishikawa}, \citenamefont {Matsufuru}, \citenamefont {Onogi}, \citenamefont
  {Yamada},\ and\ \citenamefont {Hashimoto}}]{Ishikawa:1997sx}%
  \BibitemOpen
  \bibfield  {author} {\bibinfo {author} {\bibfnamefont {K.-I.}\ \bibnamefont
  {Ishikawa}}, \bibinfo {author} {\bibfnamefont {H.}~\bibnamefont {Matsufuru}},
  \bibinfo {author} {\bibfnamefont {T.}~\bibnamefont {Onogi}}, \bibinfo
  {author} {\bibfnamefont {N.}~\bibnamefont {Yamada}}, \ and\ \bibinfo {author}
  {\bibfnamefont {S.}~\bibnamefont {Hashimoto}},\ }\href {\doibase
  10.1103/PhysRevD.56.7028} {\bibfield  {journal} {\bibinfo  {journal} {Phys.
  Rev. D}\ }\textbf {\bibinfo {volume} {56}},\ \bibinfo {pages} {7028}
  (\bibinfo {year} {1997})},\ \Eprint {http://arxiv.org/abs/hep-lat/9706008}
  {arXiv:hep-lat/9706008} \BibitemShut {NoStop}%
\bibitem [{\citenamefont {Aoki}\ \emph {et~al.}(2009)\citenamefont {Aoki} \emph
  {et~al.}}]{PACS-CS:2008bkb}%
  \BibitemOpen
  \bibfield  {author} {\bibinfo {author} {\bibfnamefont {S.}~\bibnamefont
  {Aoki}} \emph {et~al.} (\bibinfo {collaboration} {PACS-CS}),\ }\href
  {\doibase 10.1103/PhysRevD.79.034503} {\bibfield  {journal} {\bibinfo
  {journal} {Phys. Rev. D}\ }\textbf {\bibinfo {volume} {79}},\ \bibinfo
  {pages} {034503} (\bibinfo {year} {2009})},\ \Eprint
  {http://arxiv.org/abs/0807.1661} {arXiv:0807.1661 [hep-lat]} \BibitemShut
  {NoStop}%
\bibitem [{\citenamefont {Aoki}\ \emph {et~al.}(2020)\citenamefont {Aoki},
  \citenamefont {Iritani},\ and\ \citenamefont {Yazaki}}]{Aoki:2019gqt}%
  \BibitemOpen
  \bibfield  {author} {\bibinfo {author} {\bibfnamefont {S.}~\bibnamefont
  {Aoki}}, \bibinfo {author} {\bibfnamefont {T.}~\bibnamefont {Iritani}}, \
  and\ \bibinfo {author} {\bibfnamefont {K.}~\bibnamefont {Yazaki}},\ }\href
  {\doibase 10.1093/ptep/ptz166} {\bibfield  {journal} {\bibinfo  {journal}
  {PTEP}\ }\textbf {\bibinfo {volume} {2020}},\ \bibinfo {pages} {023B03}
  (\bibinfo {year} {2020})},\ \Eprint {http://arxiv.org/abs/1909.00656}
  {arXiv:1909.00656 [hep-lat]} \BibitemShut {NoStop}%
\bibitem [{\citenamefont {Haftel}\ and\ \citenamefont
  {Tabakin}(1970)}]{Haftel:1970zz}%
  \BibitemOpen
  \bibfield  {author} {\bibinfo {author} {\bibfnamefont {M.~I.}\ \bibnamefont
  {Haftel}}\ and\ \bibinfo {author} {\bibfnamefont {F.}~\bibnamefont
  {Tabakin}},\ }\href {\doibase 10.1016/0375-9474(70)90047-3} {\bibfield
  {journal} {\bibinfo  {journal} {Nucl. Phys. A}\ }\textbf {\bibinfo {volume}
  {158}},\ \bibinfo {pages} {1} (\bibinfo {year} {1970})}\BibitemShut {NoStop}%
\bibitem [{\citenamefont {Iritani}\ \emph {et~al.}(2017)\citenamefont
  {Iritani}, \citenamefont {Aoki}, \citenamefont {Doi}, \citenamefont
  {Hatsuda}, \citenamefont {Ikeda}, \citenamefont {Inoue}, \citenamefont
  {Ishii}, \citenamefont {Nemura},\ and\ \citenamefont
  {Sasaki}}]{Iritani:2017rlk}%
  \BibitemOpen
  \bibfield  {author} {\bibinfo {author} {\bibfnamefont {T.}~\bibnamefont
  {Iritani}}, \bibinfo {author} {\bibfnamefont {S.}~\bibnamefont {Aoki}},
  \bibinfo {author} {\bibfnamefont {T.}~\bibnamefont {Doi}}, \bibinfo {author}
  {\bibfnamefont {T.}~\bibnamefont {Hatsuda}}, \bibinfo {author} {\bibfnamefont
  {Y.}~\bibnamefont {Ikeda}}, \bibinfo {author} {\bibfnamefont
  {T.}~\bibnamefont {Inoue}}, \bibinfo {author} {\bibfnamefont
  {N.}~\bibnamefont {Ishii}}, \bibinfo {author} {\bibfnamefont
  {H.}~\bibnamefont {Nemura}}, \ and\ \bibinfo {author} {\bibfnamefont
  {K.}~\bibnamefont {Sasaki}},\ }\href {\doibase 10.1103/PhysRevD.96.034521}
  {\bibfield  {journal} {\bibinfo  {journal} {Phys. Rev. D}\ }\textbf {\bibinfo
  {volume} {96}},\ \bibinfo {pages} {034521} (\bibinfo {year} {2017})},\
  \Eprint {http://arxiv.org/abs/1703.07210} {arXiv:1703.07210 [hep-lat]}
  \BibitemShut {NoStop}%
\bibitem [{\citenamefont {Kamimura}(1988)}]{Kamimura:1988zz}%
  \BibitemOpen
  \bibfield  {author} {\bibinfo {author} {\bibfnamefont {M.}~\bibnamefont
  {Kamimura}},\ }\href {\doibase 10.1103/PhysRevA.38.621} {\bibfield  {journal}
  {\bibinfo  {journal} {Phys. Rev. A}\ }\textbf {\bibinfo {volume} {38}},\
  \bibinfo {pages} {621} (\bibinfo {year} {1988})}\BibitemShut {NoStop}%
\bibitem [{\citenamefont {Hashimoto}\ \emph {et~al.}(1998)\citenamefont
  {Hashimoto}, \citenamefont {Ishikawa}, \citenamefont {Matsufuru},
  \citenamefont {Onogi},\ and\ \citenamefont {Yamada}}]{Hashimoto:1997sr}%
  \BibitemOpen
  \bibfield  {author} {\bibinfo {author} {\bibfnamefont {S.}~\bibnamefont
  {Hashimoto}}, \bibinfo {author} {\bibfnamefont {K.-I.}\ \bibnamefont
  {Ishikawa}}, \bibinfo {author} {\bibfnamefont {H.}~\bibnamefont {Matsufuru}},
  \bibinfo {author} {\bibfnamefont {T.}~\bibnamefont {Onogi}}, \ and\ \bibinfo
  {author} {\bibfnamefont {N.}~\bibnamefont {Yamada}},\ }\href {\doibase
  10.1103/PhysRevD.58.014502} {\bibfield  {journal} {\bibinfo  {journal} {Phys.
  Rev. D}\ }\textbf {\bibinfo {volume} {58}},\ \bibinfo {pages} {014502}
  (\bibinfo {year} {1998})},\ \Eprint {http://arxiv.org/abs/hep-lat/9711031}
  {arXiv:hep-lat/9711031} \BibitemShut {NoStop}%
\bibitem [{\citenamefont {Amagasa}\ \emph {et~al.}(2015)\citenamefont {Amagasa}
  \emph {et~al.}}]{Amagasa:2015zwb}%
  \BibitemOpen
  \bibfield  {author} {\bibinfo {author} {\bibfnamefont {T.}~\bibnamefont
  {Amagasa}} \emph {et~al.},\ }\href {\doibase 10.1088/1742-6596/664/4/042058}
  {\bibfield  {journal} {\bibinfo  {journal} {J. Phys. Conf. Ser.}\ }\textbf
  {\bibinfo {volume} {664}},\ \bibinfo {pages} {042058} (\bibinfo {year}
  {2015})}\BibitemShut {NoStop}%
\bibitem [{\citenamefont {Ueda}\ \emph {et~al.}(2014)\citenamefont {Ueda},
  \citenamefont {Aoki}, \citenamefont {Aoyama}, \citenamefont {Kanaya},
  \citenamefont {Matsufuru}, \citenamefont {Motoki}, \citenamefont {Namekawa},
  \citenamefont {Nemura}, \citenamefont {Taniguchi},\ and\ \citenamefont
  {Ukita}}]{Ueda:2014rya}%
  \BibitemOpen
  \bibfield  {author} {\bibinfo {author} {\bibfnamefont {S.}~\bibnamefont
  {Ueda}}, \bibinfo {author} {\bibfnamefont {S.}~\bibnamefont {Aoki}}, \bibinfo
  {author} {\bibfnamefont {T.}~\bibnamefont {Aoyama}}, \bibinfo {author}
  {\bibfnamefont {K.}~\bibnamefont {Kanaya}}, \bibinfo {author} {\bibfnamefont
  {H.}~\bibnamefont {Matsufuru}}, \bibinfo {author} {\bibfnamefont
  {S.}~\bibnamefont {Motoki}}, \bibinfo {author} {\bibfnamefont
  {Y.}~\bibnamefont {Namekawa}}, \bibinfo {author} {\bibfnamefont
  {H.}~\bibnamefont {Nemura}}, \bibinfo {author} {\bibfnamefont
  {Y.}~\bibnamefont {Taniguchi}}, \ and\ \bibinfo {author} {\bibfnamefont
  {N.}~\bibnamefont {Ukita}},\ }\href {\doibase 10.1088/1742-6596/523/1/012046}
  {\bibfield  {journal} {\bibinfo  {journal} {J. Phys. Conf. Ser.}\ }\textbf
  {\bibinfo {volume} {523}},\ \bibinfo {pages} {012046} (\bibinfo {year}
  {2014})}\BibitemShut {NoStop}%
\bibitem [{\citenamefont {Vogt}\ and\ \citenamefont
  {Schr\"ock}(2015)}]{Vogt:2014hea}%
  \BibitemOpen
  \bibfield  {author} {\bibinfo {author} {\bibfnamefont {H.}~\bibnamefont
  {Vogt}}\ and\ \bibinfo {author} {\bibfnamefont {M.}~\bibnamefont
  {Schr\"ock}},\ }in\ \href {\doibase 10.3204/DESY-PROC-2014-05/31} {\emph
  {\bibinfo {booktitle} {{GPU Computing in High-Energy Physics}}}}\ (\bibinfo
  {year} {2015})\ pp.\ \bibinfo {pages} {175--180},\ \Eprint
  {http://arxiv.org/abs/1412.3655} {arXiv:1412.3655 [hep-lat]} \BibitemShut
  {NoStop}%
\end{thebibliography}%
\end{document}